\documentclass[aps,amsmath,amssymb,twocolumn,superscriptaddress,10pt]{revtex4-1}
\usepackage[T1]{fontenc}
\usepackage{graphicx}  
\usepackage[colorlinks=true,linkcolor=blue,citecolor=red,urlcolor=magenta]{hyperref}
\usepackage[dvipsnames]{xcolor}
\usepackage{charter}

\begin{document}   
  
\title{Geometry-induced entanglement in a mass-imbalanced few-fermion system}
\author{Damian W{\l}odzy\'nski}
\affiliation{Institute of Physics, Polish Academy of Sciences, Aleja Lotnik\'ow 32/46, PL-02668 Warsaw, Poland} 
\author{Daniel P{\c e}cak}
\affiliation{Faculty of Physics, Warsaw University of Technology, Ulica Koszykowa 75, PL-00662 Warsaw, Poland}
\author{Tomasz Sowi\'nski} 
\affiliation{Institute of Physics, Polish Academy of Sciences, Aleja Lotnik\'ow 32/46, PL-02668 Warsaw, Poland} 
\date{\today} 

\begin{abstract} 
Many-body systems undergoing quantum phase transitions reveal substantial growth of non-classical correlations between different parties of the system. This behavior is manifested by characteristic divergences of the von Neumann entropy. Here we show, that very similar features may be observed in one-dimensional systems of a few strongly interacting atoms when the structural transitions between different spatial orderings are driven by a varying shape of an external potential. When the appropriate adaptation of the finite-size scaling approach is performed in the vicinity of the transition point, few-fermion systems display a characteristic power-law invariance of divergent quantities.
\end{abstract} 
\maketitle  

\section{Introduction}
Since the famous paper by Bell \cite{Bellpaper1964} and its subsequent experimental confirmation \cite{aspect1982experimental} properties of non-classical correlations (quantum entanglement) and their existence in different quantum systems have been studied extensively. From the theoretical point of view, the importance of entanglement comes mainly from the fact that it is recognized as one of the key resources for efficient quantum computations and quantum information processing \cite{nielsen2002quantum}. Experimentally, due to existing experimental possibilities, these studies are mainly focused on properties of entangled photons \cite{2018FlaminiRPP}, ultra-cold atomic systems \cite{2008BlochNature}, or correlated spintronic systems \cite{2016LaflorenciePhysRep}. Properties of  quantum entanglement have also fundamental importance when interacting many-body systems close to quantum phase transitions are considered \cite{Sachdev2011QPT}. Almost 20 years ago it has been shown that non-classical correlations between different parties of a many-body system undergoing quantum transition rapidly change, {\it i.e.}, structural changes of the many-body state in the vicinity of the transition point are accompanied by characteristic divergences of the von Neumann entropy quantifying quantum entanglement~\cite{osterloh2002scaling,vidal2003entanglement,miranowicz2002generation,wu2004quantum,liu2013quantum,ma2009fisher}. This observation was recently utilized to produce entanglement on-demand in a deterministic way by driving the system through the transition~\cite{luo2017deterministic}. Not only coherent engineering and control of entanglement in the systems encounter difficulties but also direct methods of measuring many-body entanglement are usually very complex and utilize a large number of resources~\cite{daley2012measuring,lanting2014entanglement,islam2015measuring,fukuhara2015spatially}. However, some indirect ways to determine multi-particle entanglement through susceptibilities are theoretically proposed~\cite{2005WiesniakNJP,hauke2016measuring}.

Quantum transitions are typically understood as divergent changes in the ground-state of many-body systems in the thermodynamic limit, {\it i.e.}, when the system's size (number of particles, spatial width, {\it etc.}) tends to infinity along with intensive parameters keeping fixed. A typical example is an infinite chain of locally coupled spins in a transverse magnetic field. In the vicinity of the transition point, any finite-size chain behaves analytically but some quantities change more rapidly for longer chains. Appropriate scaling shows that in the limit of the infinite size of the system these quantities are exponentially divergent with well established critical exponents. The microscopic explanation of such behavior is served in the framework of the renormalization group theory \cite{1974FisherRMP}. All it means that by studying properties of finite-size chains with growing sizes, one performs appropriate extrapolation of obtained results and tries to establish appropriate relations for the system being in the thermodynamic limit.  

In fact, the limit in which critical behavior of the many-body system is manifested by divergent quantities is related to the number of accessible microscopic configurations rather than its spatial sizes or number of particles. It means that the critical behavior can be also considered in systems containing a finite number of particles allowed to occupy an infinite number of configurations. For example, as recently argued in \cite{2016PecakPRA}, strongly interacting systems of a few ultra-cold fermions of different mass confined in one-dimensional traps may undergo critical transitions when external trapping is adiabatically changed. In these cases, the thermodynamic limit (infinite number of accessible configurations) is achieved by the limit of infinite repulsions. Simply, in this limit, an infinite number of fundamentally different Fock states essentially contribute to the ground state of the system. As shown in \cite{2016PecakPRA}, in this limit the ground state undergoes a transition between different spatial distributions of components. This behavior is manifested by divergences of the second moment of the magnetization (the difference between density distributions of the heavy and light component).  

In this paper, we aim to shed some additional light on the mentioned transition in few-fermion systems by analyzing potential divergences of non-classical correlations. In this way, we want to draw a much closer analogy between critical transitions in systems containing a finite number of particles (but having infinitely many configurations accessible) with standard quantum phase transitions occurring in systems of infinite sizes. Since few-fermion systems of equal mass atoms have been engineered almost for a decade~\cite{2011SerwaneScience,2013WenzScience,2013ZurnPRL} and there are experimental setups where mass-imbalanced fermionic mixtures with a large number of particles are realized~\cite{2010TieckePRL,2011NaikEPJD,2015CetinaPRL,cetina2016ultrafast,Grimm2018DyK}, we believe that the path of exploration proposed here may be important when the next generation of experiments of {\it mass-imbalanced few-fermion mixtures} will be performed \cite{2019SowinskiRPP,2019MistakidisNJP}.

Our work is organized as follows. In Sec.~\ref{Sec2} we describe the system under consideration. In Sec.~\ref{Sec3}, to make a whole story as clear as possible, we shortly recall previous results on critical transition in few-fermion systems. In Sec.~\ref{Sec4} and Sec.~\ref{Sec5}, focusing on particularly chosen examples of Li-K fermionic mixtures, we analyze the critical transition in terms of the single-particle and the inter-component entanglement entropies, respectively. Then in Sec.~\ref{Sec6}, we explain that the discussed critical transition and the behavior of entanglement entropies in its vicinity are a very generic for few-fermion mixtures. Finally, we conclude in Sec.~\ref{Sec7}. At this point, we want also to emphasize that all results presented in this work should be considered together with accompanying supplementary material \cite{Supplement}, where results complementary to the results presented in the main text are displayed. Just for clarity of the discussion, we move these additional results out of the main text.

\section{The system studied} \label{Sec2}
In this paper we study the ground-state properties of a two-component mixture of a few ultra-cold fermions of mass $m_\sigma$ ($\sigma\in\{A,B\}$ indicates the component) confined in a one-dimensional trap with varying shape. The Hamiltonian of the system in the second quantization formalism has a form
\begin{align} \label{Hamiltonian}
\hat{\cal H} &= \sum_{\sigma\in\{A,B\}}\int\!\mathrm{d}x\,\hat{\Psi}^\dagger_\sigma(x)\left[-\frac{\hbar^2}{2m_\sigma}\frac{\mathrm{d}^2}{\mathrm{d}x^2}+V_\sigma(x)\right]\hat\Psi_\sigma(x) \nonumber \\
&+ g\int\!\mathrm{d}x\,\hat{\Psi}^\dagger_A(x)\hat{\Psi}^\dagger_B(x)\hat{\Psi}_B(x)\hat{\Psi}_A(x),
\end{align}
where $g$ is the effective interaction strength between opposite component particles and $\hat\Psi_\sigma(x)$ is a fermionic field operator annihilating a particle from the component $\sigma$ at position $x$. It obeys standard fermionic anti-commutation relations
\begin{subequations}
\begin{align}
\{\hat\Psi_\sigma(x),\hat\Psi^\dagger_{\sigma'}(x')\}&=\delta_{\sigma\sigma'}\delta(x-x'), \\
\{\hat\Psi_\sigma(x),\hat\Psi_{\sigma'}(x')\}&=0.
\end{align}
\end{subequations}
Note that the Hamiltonian (\ref{Hamiltonian}) commutes independently with operators of number of atoms $\hat{N}_\sigma=\int\!\mathrm{d}x\,\hat\Psi^\dagger_\sigma(x)\hat\Psi_\sigma(x)$. Therefore, properties of the ground state can be studied independently for given numbers of atoms in both components $N_A$ and $N_B$. Here, to make a whole analysis as clear as possible, we focus on the problem of $N_A+N_B=4$ particles with different distributions between the components. However, generalization to higher number of particles is straightforward. As an example, the case of $N_A+N_B=6$ particles is briefly discussed in Sec.~\ref{Sec6}.

To find the many-body ground state $|\mathtt{G}\rangle$ of the Hamiltonian \eqref{Hamiltonian} we perform straightforward diagonalization of its matrix representation in the Fock basis of the non-interacting system. First, for a given external confinement $V_\sigma(x)$, we numerically solve the corresponding single-particle eigenproblems for each component $\sigma$
\begin{equation}
\left[-\frac{\hbar^2}{2m_\sigma}\frac{\mathrm{d}^2}{\mathrm{d}x^2}+V_\sigma(x)\right]\varphi^{(\sigma)}_i(x) = E^{(\sigma)}_i\varphi^{(\sigma)}_i(x)
\end{equation}
and find single-particle eigenenergies $E^{(\sigma)}_i$ and eigenfunctions $\varphi^{(\sigma)}_i(x)$. Then, for a fixed number of particles $N_A$ and $N_B$, the Fock basis is constructed from $K$ the lowest orbitals. Each Fock basis element (a Fock state) is a simple product of two Slater determinants of $N_A$ ($N_B$) orbitals corresponding to the many-body state of $A$-component ($B$-component) fermions. The dimension of the resulting Hilbert space spanned by this Fock basis is ${\cal D}={K\choose N_A}\cdot{K\choose N_B}$. Additionally, since the Hamiltonian \eqref{Hamiltonian} conserves the parity of many-body states \cite{1998HaugsetPRA}, we remove from the basis the states which indisputably cannot contribute to the many-body ground state. Importantly, for a larger number of particles, to increase numerical efficiency, we remove also some irrelevant high-energy Fock states from the basis according to the prescription given in \cite{2019ChrostowskiAPPA}. Finally, for a given interaction strength $g$, all matrix elements of the full many-body Hamiltonian \eqref{Hamiltonian} in the prepared basis are calculated and the resulting matrix is numerically diagonalized. Since we are interested only in the properties of the many-body ground state and the resulting matrix is essentially sparse, we use the Arnoldi method \cite{ArnoldiBook} to obtain only the eigenvector corresponding to the smallest energy. A numerical convergence is obtained by increasing the cut-off $K$ and checking the ground-state fidelity. We found that for considered the number of particles and interaction strengths, the optimal cut-off varies from $K=14$ to $K=30$ depending on the shape of the trap and number of particles.  

\section{The transition} \label{Sec3}
It was argued recently in \cite{2015LoftEPJD,2016PecakNJP,2016DehkharghaniJPhysB}, that in the case of a different mass of particles ($m_A\neq m_B$) and whenever the interaction strength $g$ is strong enough, one observes a characteristic spatial separation of components which is manifested in properties of the many-body ground state of the system $|\mathtt{G}\rangle$. One of the most visible footprints of this separation is present in the density profiles (normalized to the number of particles) of the components
\begin{equation} \label{DensityProfile}
n_\sigma(x) = \langle \mathtt{G}|\hat\Psi^\dagger_\sigma(x)\hat\Psi_\sigma(x)|\mathtt{G}\rangle.
\end{equation}
Namely, the profile of one of the components is split into two parts and it is push-out from the center of the trap, while the second component's profile remains localized in the center. Interestingly, depending on details of the external confinement, the separation is induced in the component of lighter or heavier particles. For example, in the case of harmonic confinement, the lighter component is pushed out from the center \cite{2015LoftEPJD,2016PecakNJP,2016DehkharghaniJPhysB}. Contrary, in the case of flat box potential, lighter particles remain in the center while the heavier component is separated. Consequently, by an adiabatic change of the shape one observes the transition between different spatial orderings which are nicely signaled by a rapid change of the second moment of the magnetization, ${\cal M}(x)=n_A(x)-n_B(x)$ \cite{2016PecakPRA}. It was shown, that the transition becomes more rapid when interaction strength is enhanced. By applying appropriate scaling in the vicinity of the transition point it was argued that in the limit of infinite repulsions the transition has properties similar to the second-order phase transitions. It is worth noticing however that the similarity to standard phase transitions is highly indirect since the system studied contains always a finite number of particles and its spatial size is far from the macroscopic. In the case studied, the thermodynamic limit is mimicked by the limit of infinite interactions rather than the size of the system. All details of this analogy are explained in \cite{2016PecakPRA}.

\begin{figure}
\centering
\includegraphics[width=\linewidth]{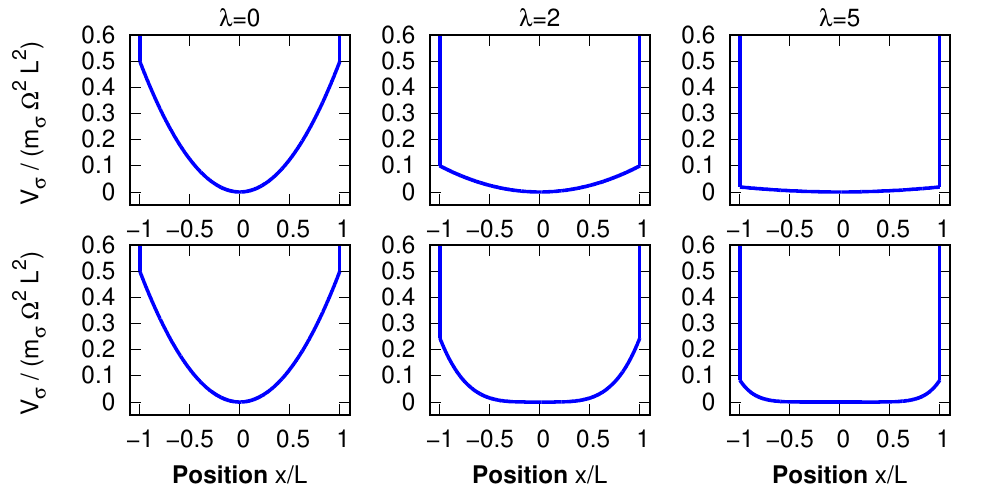}
\caption{Shape of an external potential \eqref{Potencjal} for different values of parameter $\lambda$ and two different transition functions $f_1(\lambda)$ (upper row) and $f_2(\lambda)$ (bottom row). For $\lambda=0$ harmonic confinement (cropped at $|x|=L$) with frequency $\Omega$ is restored. In the limit $\lambda\rightarrow\infty$ the trap is equivalent to the box of the length $L$.\label{Fig1}}
\end{figure}

Here, we want to significantly extend previous results and perform detailed quantitative studies of this transition from the quantum correlations point of view. We also want to answer the question if the properties of the transition qualitatively depend on the protocol changing the shape. To address these two issues, in the following we consider a general scenario of switching the confinement from the harmonic trap of frequency $\Omega$ to the flat box trap of width $2L$ as following
\begin{equation} \label{Potencjal}
V_\sigma(x) = \left\{
\begin{array}{cl}
f(\lambda)\frac{m_\sigma\Omega^2}{2}x^2, & |x|<L, \\
\infty, & |x|\geq L
\end{array}
\right.
\end{equation}
where the function $f(\lambda)$ (in general position-dependent) is chosen in such a way that it smoothly and monotonically changes (for each position $x$ independently) from $1$ to $0$ when the dimensionless parameter $\lambda$ is tuned from $0$ to $\infty$. In the following, we consider two different complementary scenarios of such switching. In the first, similarly as was done in \cite{2016PecakPRA}, we simply assume that the function $f(\lambda)$ does not depend on particles' positions and it has a form $f_1(\lambda)=(1+\lambda^2)^{-1}$. It means that for any $\lambda$ the shape of the confinement remains parabolic in the middle of the trap and only its frequency is changed. In the second scenario the harmonic part is turned off by the substantial change of its shape and the function $f(\lambda)$ depends on positions. Motivated with recent experiments in which the flatness of the trap is bounded by a power function \cite{2010vanEsJPhysB,2013GauntPRL,2015ChomazNatComm,2017MukherjeePRL,2018HueckPRL}, we model this protocol assuming that $f_2(\lambda)=|x/L_0|^\lambda$, with some well defined $L_0>L$. In fact, $L_0$ determines how fast the confinement changes between limiting cases. To build some intuition, in Fig.~\ref{Fig1}, we display shapes of the trap for several different values of $\lambda$ for both scenarios considered. In Fig.~S1 of the supplementary material \cite{Supplement} we present corresponding single-particle spectra as function of parameter $\lambda$.

For convenience, in further considerations, we express all quantities in natural units of the $A$-component harmonic oscillator, {\it i.e.}, we express all energies, lengths, and momenta in $\hbar\Omega$, $\sqrt{\hbar/m_A\Omega}$, and $\sqrt{\hbar m_A\Omega}$ respectively. Thus, the coupling constant is scaled by a factor of $\sqrt{\hbar^3\Omega/m_A}$. In these units dimensionless masses are $\mu_A=1$ and $\mu_B=m_B/m_A$. To keep correspondence with the previous results presented in \cite{2016PecakPRA} we set $L=3.5$ and $L_0=5$ in these units.

\section{Single-particle correlations} \label{Sec4}
\begin{figure}[t]
\centering
\includegraphics[width=1.05\linewidth]{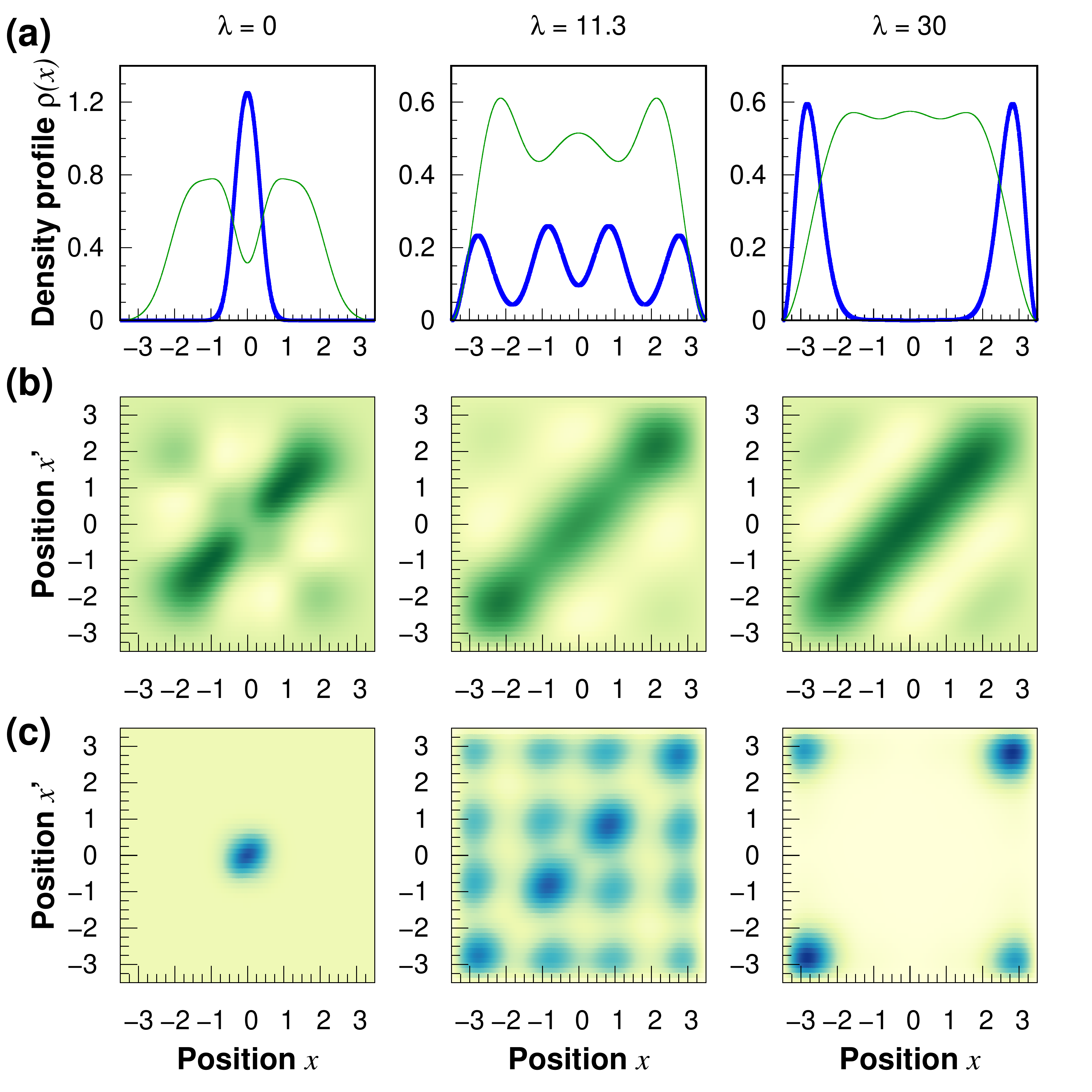}
\caption{Structural transition in the ground state of the system of $N_A=3$ and $N_B=1$ interacting fermions from a single-particle point of view ($m_B/m_A=40/6$). Successive columns (from left to right) correspond to different external traps, from the harmonic oscillator ($\lambda=0$) to the flat box ($\lambda\rightarrow\infty$). (a) Single-particle density profile for heavier (thick blue) and lighter (thin green) component depending on the shape of the external trap. (b-c) The single-particle density matrix of the lighter and heavier component, respectively. Similar results for other distributions of particles between components are given in Fig.~S2 in the supplementary material \cite{Supplement}. \label{Fig2}}
\end{figure}
To give a better understanding of the properties of the system for different confinements $\lambda$ let us first focus on the ground-state properties of a system with $N_A=3$ and $N_B=1$ particles confined in the trap described by the function $f_1(\lambda)$. In Fig.~\ref{Fig2} we present different single-particle properties for quite strong repulsions $g=5$ and reasonable mass ratio $\mu_B= 40/6$ corresponding to Li-K fermionic mixture. In Fig.~\ref{Fig2}a single-particle density profiles $n_\sigma(x)$ for the lighter and heavier component are presented (thin green and thick blue lines, respectively). As it is seen, depending on the shape of the trap, the lighter or the heavier component is split and pushed out from the center. It is clearly seen that in the harmonic potential ($\lambda=0$) the separation is present in the light atoms component. Contrary, in the trap shape close to the uniform box potential ($\lambda\rightarrow\infty$), the separation is induced in heavier atoms component. Note, that for some particular intermediate shape of the confinement ($\lambda_0\approx 11.3$) a specific transition between these two scenarios is present and characteristic oscillations in the density profiles appear. This transition in the structure of the many-body ground state induced by an adiabatic change of the external potential was studied recently in \cite{2016PecakPRA}. 

\begin{figure}[t]
\centering
\includegraphics[width=\linewidth]{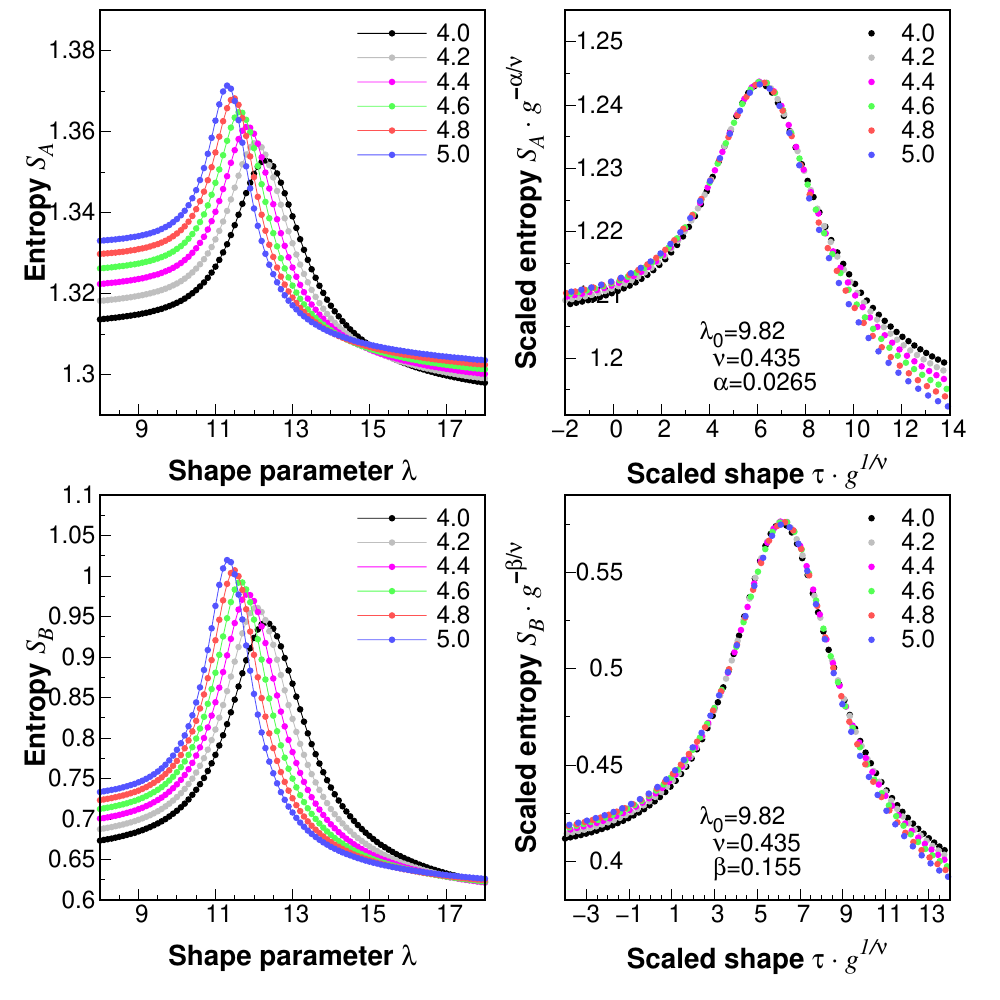}\caption{The single-particle von Neumann entropies ${\cal S}_A$ and ${\cal S}_B$ (top and bottom row respectively) defined by \eqref{SingleEntropy} as a function of the shape of external trapping for the same system as in Fig.~\ref{Fig2}. Note that with appropriate scaling results obtained for different interactions collapse to the same universal curve (right panel). See the main text for details. For other distributions of particles see Fig.~S3 in the supplementary material \cite{Supplement}.\label{Fig3}}
\end{figure}

Much better understanding of this transition can be given when, instead of the single-particle density profile $n_\sigma(x)$, one considers a whole single-particle reduced density matrix defined straightforwardly as
\begin{equation} \label{DensMat1}
\rho_\sigma(x,x') = \frac{1}{N_\sigma}\langle\mathtt{G}|\hat\Psi^\dagger_\sigma(x)\hat\Psi_\sigma(x')|\mathtt{G}\rangle.
\end{equation}
Note that here, in contrast to \eqref{DensityProfile}, we normalize the density matrix to unity.
This quantity is sufficient when any single-particle measurement is considered since it encodes not only all diagonal but also off-diagonal single-particle correlations in any representation. In Fig.~\ref{Fig2}b and Fig.~\ref{Fig2}c we display the single-particle density matrix for both components for external confinement parameters $\lambda$ corresponding to the density profiles displayed in Fig.~\ref{Fig2}a. Substantial changes of the single-particle density matrices are clearly visible for both components. Especially, in the case of the component containing $N_B=1$ particle (Fig.~\ref{Fig2}c) the transition between limiting regimes is quite spectacular and it is manifested by strong off-diagonal correlations (middle plot). Moreover, even in the case of flat box potential ($\lambda\rightarrow\infty$), when spatial separation of both components is clearly visible, the state of the single-particle component is obviously not a pure state, {\it i.e.}, any outcome of the single-particle measurement performed on this component depends also on the state of the second component. This observation leads us directly to the conclusion that both components, due to their strong interactions, are {\it entangled}. The amount of these non-classical correlations between a particle from selected component $\sigma$ and the rest of the system are encoded in the von Neumann entropies
\begin{equation} \label{SingleEntropy}
{\cal S}_\sigma = \sum_i\lambda_{\sigma i}\ln\lambda_{\sigma i},
\end{equation}
where $\lambda_{\sigma i}$ are the egienvalues of corresponding single-particle reduced density matrices \eqref{DensMat1} obtained after performing their spectral decompositions 
\begin{equation}
\rho_\sigma(x,x') = \sum_i \lambda_{\sigma i}\eta_{\sigma i}^*(x)\eta_{\sigma i}(x')
\end{equation}
with $\eta_{\sigma i}(x)$ being their natural single-particle orbitals. Since in general particles belonging to opposite components have different masses, resulting entropies are different.

In Fig.~\ref{Fig3} (left panel) we present both entropies ${\cal S}_\sigma$ as a function of the shape of the external confinement $\lambda$ for the system of $N_A=3$ and $N_B=1$ particles and different interactions. As seen, the quantum correlations of the selected particle with the rest of the system are the strongest in the vicinity of the transition between different spacial orderings. This observation directly supports previously outlined analogy between the structural transition of the system and the quantum phase transitions of other quantum systems, since it is known that standard quantum phase transitions are very often accompanied with rapid changes of inherent correlations \cite{osterloh2002scaling}. This analogy is even more evident when we consider the limit of infinite repulsions ($g\rightarrow\infty$) which (as argued in \cite{2016PecakPRA} and discussed in Sec.~\ref{Sec2}) mimics the thermodynamic limit where different thermodynamics quantities are divergent. Of course, the limit of infinite repulsions is beyond our numerical possibilities. However, to capture properties of the single-particle entanglement entropies ${\cal S}_\sigma$ in this limit, we perform {\it finite-size} scaling assuming that in the vicinity of the transition each entropy is divergent with corresponding critical exponent $\alpha$ and $\beta$, for ${\cal S}_A$ and ${\cal S}_B$ respectively. Moreover, the system poses some additional scaling invariance related to the interaction strength characterized by a common characteristic critical exponent $\nu$ \cite{2008GuPRA,2015SowinskiPRA,2016PecakPRA}. It means that for any interaction strength $g$ the entropies can be expressed in terms of the unique (for a given number of particles and mass ratio) universal function $\mathbf{S}_\sigma(\xi)$ as
\begin{subequations} \label{Scalling}
\begin{align}
{\cal S}_A(\lambda,g) &= g^{\alpha/\nu} \mathbf{S}_A(g^{1/\nu} \tau), \\
 {\cal S}_B(\lambda,g) &= g^{\beta/\nu} \mathbf{S}_B(g^{1/\nu} \tau)
\end{align}
\end{subequations}
where $\tau=(\lambda-\lambda_0)/\lambda_0$ is a normalized distance from the transition point $\lambda_0$. Indeed, our numerical analysis fully confirms that there exist exactly one set of parameters $\{\lambda_0,\alpha,\beta,\nu\}$ for which all numerically obtained data points {\it collapse} to appropriate universal curves (see right column in Fig.~\ref{Fig3}). Although the data-collapse algorithm was performed independently for both entropies ${\cal S}_A$ and ${\cal S}_B$, the resulting common parameters $\lambda_0$ and $\nu$ are (up to a numerical accuracy) exactly the same. Only individual critical exponents $\alpha$ and $\beta$ are different since they characterize divergent behavior of different quantities. This suggests that indeed, in the limit of infinite repulsions, the system undergoes a structural transition at $\lambda_0\approx 9.82$. This value is evidently shifted when compared to the corresponding value in Fig.~\ref{Fig2} obtained purely phenomenologically by an eye-detection of a change in the single-particle density profiles. This difference comes directly from the fact that the results in Fig.~\ref{Fig2} are presented for finite repulsion $g=5$ while the critical shape value $\lambda_0$ corresponds to the limit of infinite repulsions. We checked that very similar transition is present for any distribution of particles between components (see Fig.~S3 in the supplementary material \cite{Supplement}). Obviously, the position $\lambda_0$ and critical exponents $\alpha$, $\beta$, and $\nu$ crucially depend on this distribution as well as on the mass ratio $\mu$.

\section{Inter-component correlations} \label{Sec5}
\begin{figure}[t]
\centering
\includegraphics[width=\linewidth]{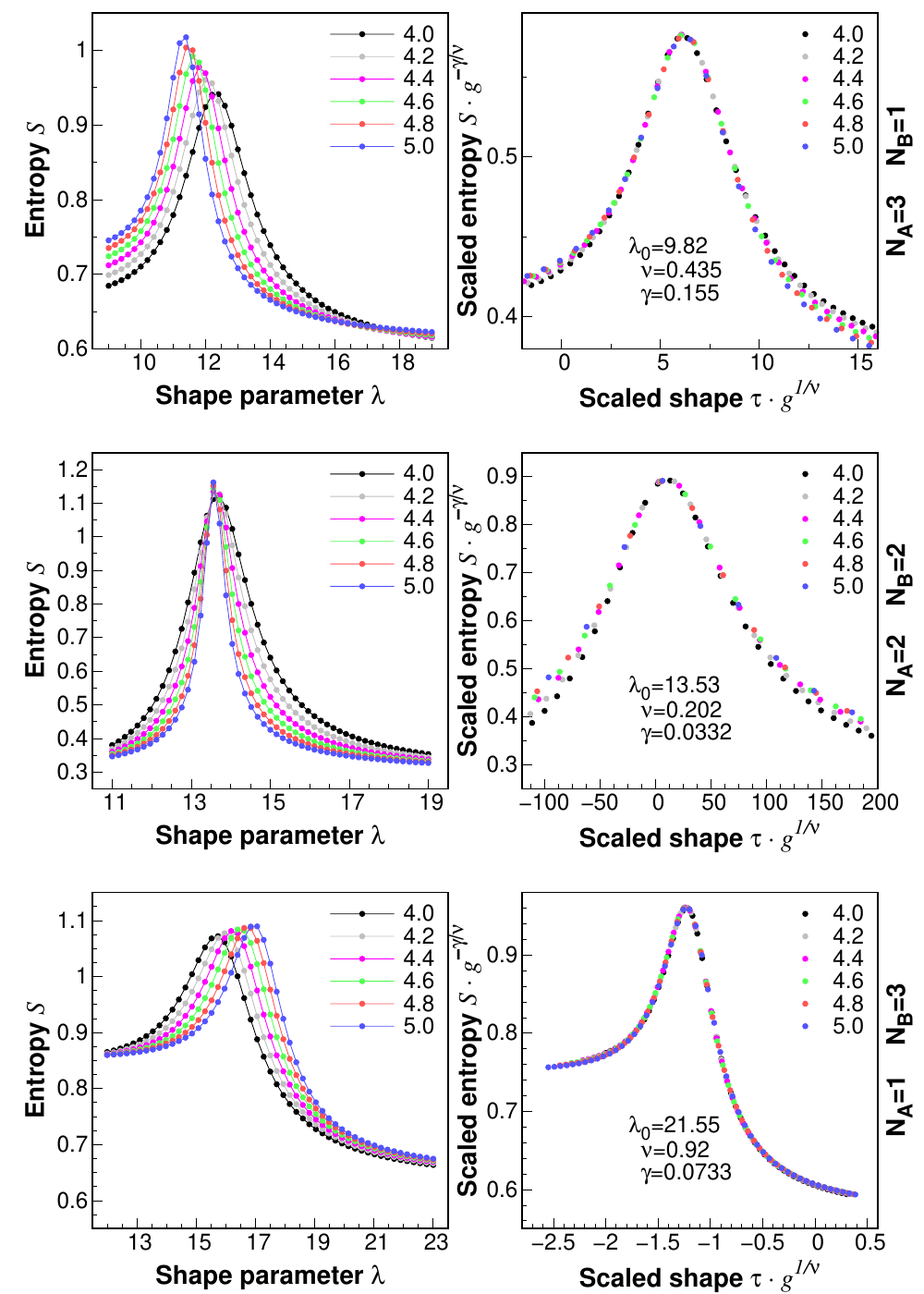}\caption{The inter-component entanglement entropy as a function of a shape of external trap $\lambda$ and interaction strength $g$ calculated for different distribution of $N=4$ particles and assumed mass ratio $m_B/m_A=40/6$. Note that in all cases appropriate scaling leads to a collapse of all data points to single universal curve. Results for other mass ratio are provided in Fig.~S4 in the supplementary material \cite{Supplement}. \label{Fig4}}
\end{figure}

In the case of $N_A=3$ and $N_B=1$ particles discussed above, the single-particle entropy ${\cal S}_B$ quantifies not only correlations between the particle and the rest of the system but in fact, it is also the entanglement entropy between components treated as a whole. This simple observation suggests that for any distribution of particles between components, probably not only the single-particle entropies are divergent in the vicinity of the structural transition but also the inter-component von Neumann entropy (quantifying entanglement between distinguishable components) has this property. To check it, we repeat all calculations for the system of $N_A=N_B=2$ with the same mass ratio $\mu=40/6$. First, we introduce the reduced density operator $\hat\Gamma_\sigma$ for a component $\sigma$ through a straightforward tracing out of all degrees of freedom of the remaining component $\sigma'$ from the projector to the many-body ground-state, $\hat{\Gamma}_\sigma=\mathrm{Tr}_{\sigma'} |\mathtt{G}\rangle\langle\mathtt{G}|$. Then, we define the inter-component von Neumann entropy as
\begin{equation} \label{CompEntropy}
{\cal S} = -\mathrm{Tr}_A\Big[\hat{\Gamma}_A \,\mathrm{ln}(\hat{\Gamma}_A)\Big]=-\mathrm{Tr}_B\Big[\hat{\Gamma}_B \,\mathrm{ln}(\hat{\Gamma}_B)\Big].
\end{equation}
It should be pointed here that, in contrast to the single-particle entanglement entropies ${\cal S}_\sigma$, the inter-component entanglement ${\cal S}$ quantifies only these correlations which are forced by mutual interactions and it vanishes in the case of the non-interacting system. Contrary, the single-particle entropies \eqref{SingleEntropy} are also sensitive to trivial correlations induced by the quantum statistics between indistinguishable fermions belonging to the same component \cite{2001SchliemannPRA}. From this point of view, both entropies (${\cal S}_\sigma$ and ${\cal S}$) give complementary information and they quantify slightly different correlations.

In Fig.~\ref{Fig4} we plot the inter-component entropy ${\cal S}$ defined by \eqref{CompEntropy} for different distributions of $N_A+N_B=4$ particles between components (left panel) and their collapse to the unique universal curves $\mathbf{S}(\xi)$ after appropriate finite-size scalings defined with the exact analogy to \eqref{Scalling} but with another critical exponent $\gamma$
\begin{equation}
{\cal S}(\lambda,g) = g^{\gamma/\nu} \mathbf{S}(g^{1/\nu} \tau)
\end{equation}
It is clearly seen, that this entropy has also required scaling invariance, {\it i.e.}, in the limit of infinite repulsions it is divergent in the vicinity of the structural transition point. It should be also underlined that values of the critical shape parameters $\lambda_0$ and the common critical exponent $\nu$ predicted with this scaling are in full agreement with those predicted from behaviors of the single-particle entanglement entropies (compare with values provided in Fig.~\ref{Fig3} and Fig.~S3 in the supplementary material). Importantly, a full consistency is confirmed by the observation that for systems with $N_A=1$ ($N_B=1$) fermions the critical exponent $\gamma$ is exactly the same as the exponent $\alpha$ ($\beta$).
\begin{figure}
\centering
\includegraphics[width=\linewidth]{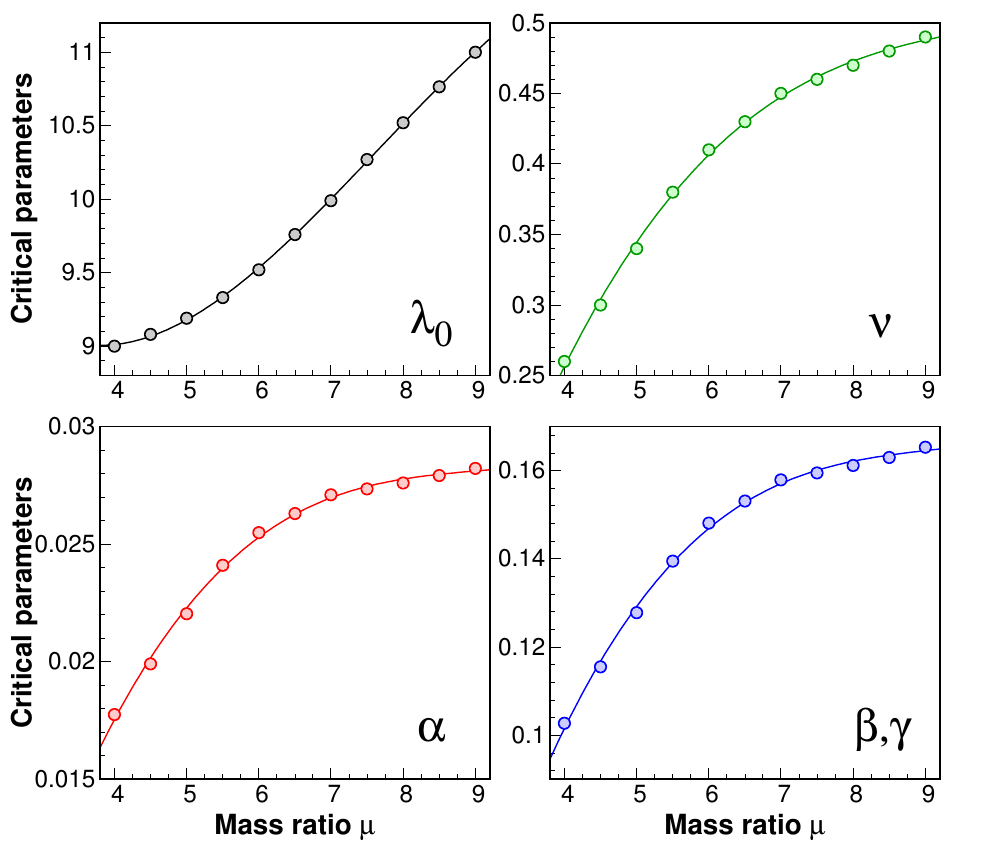}
\caption{Values of the critical parameters $\lambda_0$, $\nu$, $\alpha$, and $\beta$ as functions of the mass ratio $\mu$ for the systems with $N_A=3$, and $N_B=1$ fermions. Note that in this case the critical exponent $\gamma$ is the same as $\beta$. It is clear that all the parameters are significantly and monotonically dependent on the ratio. Results for other distributions of particles are provided in Fig.~S5 in the supplementary material \cite{Supplement}. \label{Fig5}
} 
\end{figure} 

\section{Universality of the phenomenon} \label{Sec6}
Finally, to expose that observed transition is a generic behavior of two-component fermionic mixtures in one-dimensional traps, let us shortly discuss similar systems with other parameters. First, we want to underline that the spatial separation of components is forced not only by mutual repulsions between particles as predicted in \cite{2014LindgrenNJP} but also by a mass difference of atoms belonging to opposite components \cite{2015LoftEPJD,2016PecakNJP,2016DehkharghaniJPhysB,2016PecakPRA,2017ChungPRA,2017VolosnievFBS}. It becomes obvious when the system of an equal number of particles is considered, $N_A=N_B$. Then, in the equal mass case $\mu=m_B/m_A=1$, one finds exact symmetry between components. Therefore, their spatial distributions must be exactly the same independently on the interaction strength. Therefore, any spatial separation of the density profiles cannot be noticed. It is a matter of fact, that strength of mutual repulsions needed to force the system to separate depends on the mass ratio -- larger mass ratios require lower repulsions to separate the system. In Fig.~S4 of the supplementary material \cite{Supplement} we show how the transition is reflected in the behavior of the inter-component entanglement entropy for different mass ratios. Although qualitatively situation remains always the same, quantitatively divergences near the transition point crucially depend on the mass ratio. It is clearly visible when critical parameters are displayed as functions of the mass ratio (Fig.~\ref{Fig5}). Along with increasing mass ratio the common critical exponent $\nu$, as well as all the other critical exponents $\alpha$, $\beta$, and $\gamma$ change monotonically their values. It simply means that the mass ratio $\mu$ uniquely defines a set of critical parameters, which determine the singular behavior of measurable quantities. In the language of the theory of phase transitions, it would mean that the mass ratio determines some kind of the universality class of the transition.

\begin{figure}
\centering
\includegraphics[width=\linewidth]{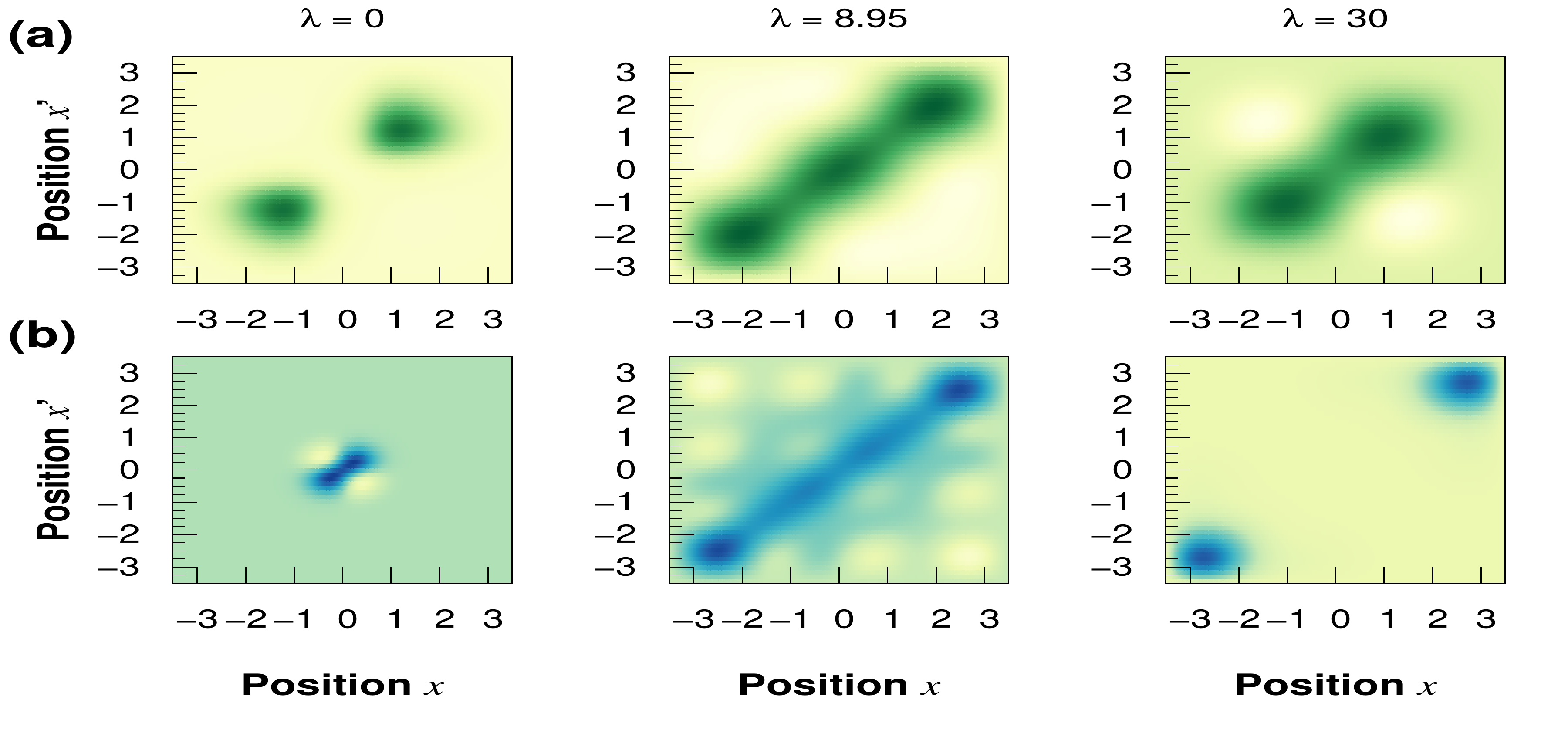}\\\includegraphics[width=\linewidth]{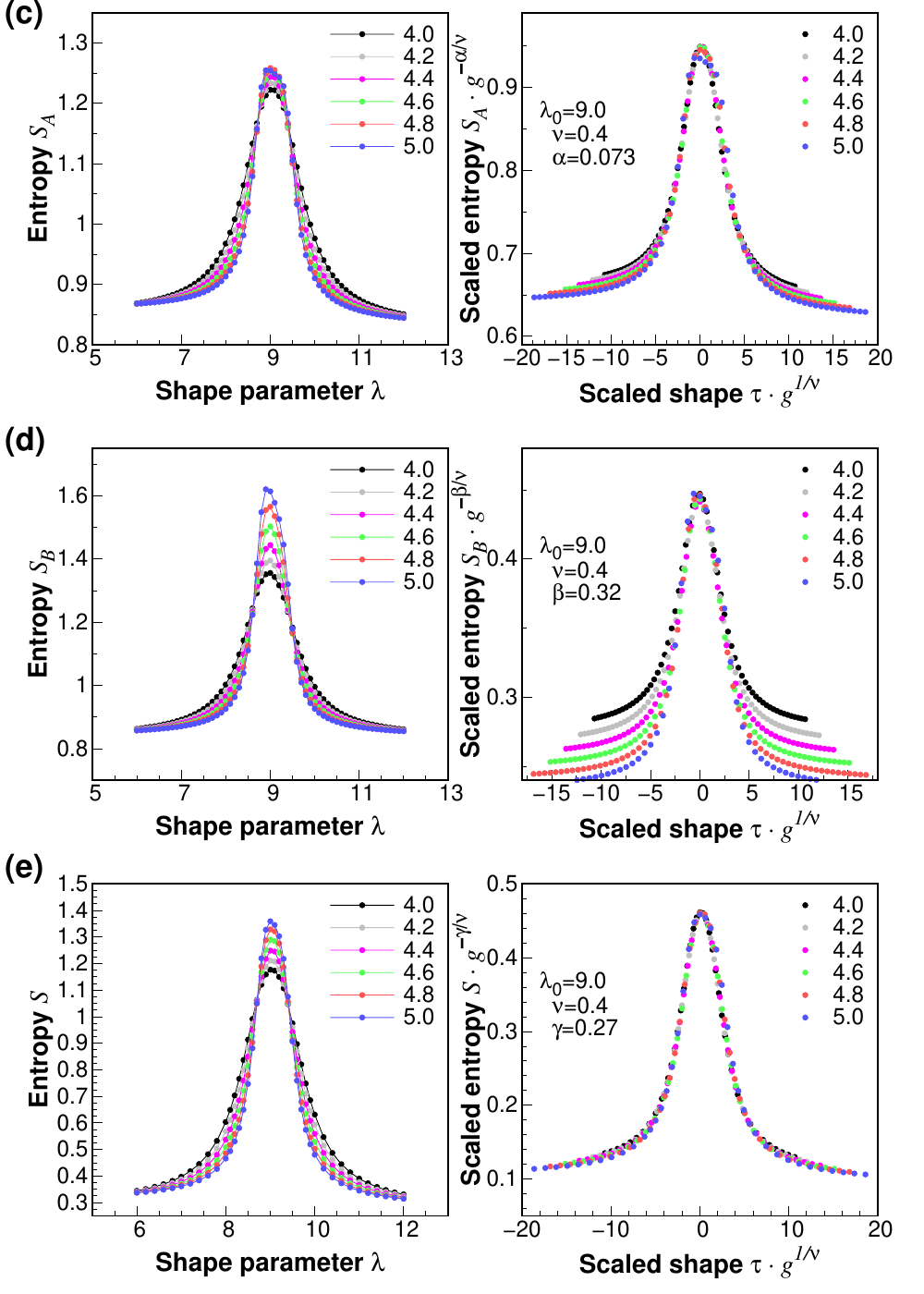}
\caption{Structural transition in the ground state of the system of $N_A=N_B=2$ fermions ($m_A/m_B=40/6$) with confinement's shape described by the function $f_2(\lambda)$. (a-b) Single-particle density matrix of the lighter and heavier component (first and second row respectively), for interaction strength $g=5$. Successive columns (from left to right) correspond to different external traps, from the harmonic oscillator ($\lambda=0$) to the flat box ($\lambda\rightarrow\infty$). (c-d) The single-particle von Neumann entropies $S_A$ and $S_B$ as a function of the shape parameter $\lambda$ and interaction strength $g$, with corresponding universal scaling (right panel). (e) Inter-component entropy $S$ and its scaling for the same system. \label{Fig6}
} 
\end{figure}

Secondly, it should be also emphasized that the structural transition in the many-body ground state is a generic behavior of one-dimensional fermionic mixtures and it is always characterized by a divergent behavior of the entanglement entropy. This behavior is insensitive to details of the protocol of switching between different shapes. As one of the examples, in Fig.~\ref{Fig6} we present corresponding results for the system of $N_A=N_B=2$ fermions when the transition between harmonic and flat potential is described with the function $f_2$ (results for other distributions between components are presented in Fig.~S6 of the supplementary material \cite{Supplement}). It is clear that for any confinement besides the critically shaped trap the ground-state structure is well established: for strong enough repulsions the heavier or the lighter component is split and pushed out from the center, while the remaining component is located in the center. At the transition point, the spatial structure is rapidly swapped and a significant increase (divergence in the limit of infinite interactions) of inter-particle and inter-component correlations (quantified with entanglement entropies) is observed.  

Finally, we want to mention that the structural transition described above is present also for the other number of particles and their distribution between components. In Fig.~S7 of the supplementary material \cite{Supplement} we present results for some other exemplary systems with $N=6$ particles. In all these cases the transition has exactly the same generic properties -- lighter or heavier component is split for two extremal confinements and the rapid change between these structures is present when the potential is adiabatically driven through the critical confinement. 
 
\section{Conclusions} \label{Sec7}
We have analyzed the separation mechanism induced by varying shape of external confinement in the ground state of a two-component mixture of a few repulsively interacting fermions of different mass from the correlations point of view. We show that the transition between different orderings manifested spectacularly as a rapid change of the component being separated is accompanied by a rapid and significant increase of the entanglement entropy between components as well as entanglement entropy between selected particle and the rest of the system. By performing appropriate finite-size scaling analysis we show that in the vicinity of the transition point the system has some scale-invariance which is very similar to the scaling of many-body systems in the thermodynamic limit. However, in the case studied the infinite size is replaced by infinite inter-component repulsions. We argue that at the transition point, in the limit of infinite repulsions, both entropies are divergent which brings us closer to a mentioned analogy with many-body systems. 

At this point we want to emphasize that although we consider here purely one-dimensional problem, the aspect of higher dimensionality could be also worth deeper consideration. There are at least two fundamental reasons for this: {\it (i)} Criticality in the vicinity of the transition point usually significantly depends on the dimensionality of the problem. {\it (ii)} Systems of higher dimensions have a typically larger capacity to store quantum correlations. In fact, the existence of the phase separation driven by the mass imbalance has been already reported for two- and three-dimensional cases \cite{2013CuiPRL}. However, the problem of criticality near the transition point for such systems remains obscure. The situation is also similar when some other than $s$-wave forms of inter-particle interactions are considered. For example, it is known that inter-particle $p$-wave scattering may lead to the phase separation of fermionic components \cite{kang2018spontaneous}, but critical properties are still not explored. From this point of view, our results may give some additional motivation for undertaking these paths of research and bring us closer to answer the following question: is the phase separation in strongly interacting systems sufficient for supporting critical behavior of correlations?

\acknowledgments
The authors thank Adam Miranowicz for his suggestions and critical comments which helped in improving the manuscript. This work was supported by (Polish) National Science Center Grants No. 2016/22/E/ST2/00555 (TS and DW) and 2017/27/B/ST2/02792 (DP). Numerical calculations were partially carried out in the Interdisciplinary Centre for Mathematical and Computational Modelling, University of Warsaw (ICM), under Computational Grant No. G75-6.

\bibliography{_BibTotal}

\begin{thebibliography}{52}%
\makeatletter
\providecommand \@ifxundefined [1]{%
 \@ifx{#1\undefined}
}%
\providecommand \@ifnum [1]{%
 \ifnum #1\expandafter \@firstoftwo
 \else \expandafter \@secondoftwo
 \fi
}%
\providecommand \@ifx [1]{%
 \ifx #1\expandafter \@firstoftwo
 \else \expandafter \@secondoftwo
 \fi
}%
\providecommand \natexlab [1]{#1}%
\providecommand \enquote  [1]{``#1''}%
\providecommand \bibnamefont  [1]{#1}%
\providecommand \bibfnamefont [1]{#1}%
\providecommand \citenamefont [1]{#1}%
\providecommand \href@noop [0]{\@secondoftwo}%
\providecommand \href [0]{\begingroup \@sanitize@url \@href}%
\providecommand \@href[1]{\@@startlink{#1}\@@href}%
\providecommand \@@href[1]{\endgroup#1\@@endlink}%
\providecommand \@sanitize@url [0]{\catcode `\\12\catcode `\$12\catcode
  `\&12\catcode `\#12\catcode `\^12\catcode `\_12\catcode `\%12\relax}%
\providecommand \@@startlink[1]{}%
\providecommand \@@endlink[0]{}%
\providecommand \url  [0]{\begingroup\@sanitize@url \@url }%
\providecommand \@url [1]{\endgroup\@href {#1}{\urlprefix }}%
\providecommand \urlprefix  [0]{URL }%
\providecommand \Eprint [0]{\href }%
\providecommand \doibase [0]{http://dx.doi.org/}%
\providecommand \selectlanguage [0]{\@gobble}%
\providecommand \bibinfo  [0]{\@secondoftwo}%
\providecommand \bibfield  [0]{\@secondoftwo}%
\providecommand \translation [1]{[#1]}%
\providecommand \BibitemOpen [0]{}%
\providecommand \bibitemStop [0]{}%
\providecommand \bibitemNoStop [0]{.\EOS\space}%
\providecommand \EOS [0]{\spacefactor3000\relax}%
\providecommand \BibitemShut  [1]{\csname bibitem#1\endcsname}%
\let\auto@bib@innerbib\@empty
\bibitem [{\citenamefont {Bell}(1964)}]{Bellpaper1964}%
  \BibitemOpen
  \bibfield  {author} {\bibinfo {author} {\bibfnamefont {J.~S.}\ \bibnamefont
  {Bell}},\ }\href {\doibase 10.1103/PhysicsPhysiqueFizika.1.195} {\bibfield
  {journal} {\bibinfo  {journal} {Physics Physique Fizika}\ }\textbf {\bibinfo
  {volume} {1}},\ \bibinfo {pages} {195} (\bibinfo {year} {1964})}\BibitemShut
  {NoStop}%
\bibitem [{\citenamefont {Aspect}\ \emph {et~al.}(1982)\citenamefont {Aspect},
  \citenamefont {Dalibard},\ and\ \citenamefont
  {Roger}}]{aspect1982experimental}%
  \BibitemOpen
  \bibfield  {author} {\bibinfo {author} {\bibfnamefont {A.}~\bibnamefont
  {Aspect}}, \bibinfo {author} {\bibfnamefont {J.}~\bibnamefont {Dalibard}}, \
  and\ \bibinfo {author} {\bibfnamefont {G.}~\bibnamefont {Roger}},\ }\href
  {\doibase 10.1103/PhysRevLett.49.1804} {\bibfield  {journal} {\bibinfo
  {journal} {Phys. Rev. Lett.}\ }\textbf {\bibinfo {volume} {49}},\ \bibinfo
  {pages} {1804} (\bibinfo {year} {1982})}\BibitemShut {NoStop}%
\bibitem [{\citenamefont {Nielsen}\ and\ \citenamefont
  {Chuang}(2000)}]{nielsen2002quantum}%
  \BibitemOpen
  \bibfield  {author} {\bibinfo {author} {\bibfnamefont {M.~A.}\ \bibnamefont
  {Nielsen}}\ and\ \bibinfo {author} {\bibfnamefont {I.}~\bibnamefont
  {Chuang}},\ }\href@noop {} {\emph {\bibinfo {title} {Quantum computation and
  quantum information}}}\ (\bibinfo  {publisher} {Cambridge University Press},\
  \bibinfo {address} {Cambridge},\ \bibinfo {year} {2000})\BibitemShut
  {NoStop}%
\bibitem [{\citenamefont {Flamini}\ \emph {et~al.}(2018)\citenamefont
  {Flamini}, \citenamefont {Spagnolo},\ and\ \citenamefont
  {Sciarrino}}]{2018FlaminiRPP}%
  \BibitemOpen
  \bibfield  {author} {\bibinfo {author} {\bibfnamefont {F.}~\bibnamefont
  {Flamini}}, \bibinfo {author} {\bibfnamefont {N.}~\bibnamefont {Spagnolo}}, \
  and\ \bibinfo {author} {\bibfnamefont {F.}~\bibnamefont {Sciarrino}},\ }\href
  {\doibase 10.1088/1361-6633/aad5b2} {\bibfield  {journal} {\bibinfo
  {journal} {Reports on Progress in Physics}\ }\textbf {\bibinfo {volume}
  {82}},\ \bibinfo {pages} {016001} (\bibinfo {year} {2018})}\BibitemShut
  {NoStop}%
\bibitem [{\citenamefont {Bloch}(2008)}]{2008BlochNature}%
  \BibitemOpen
  \bibfield  {author} {\bibinfo {author} {\bibfnamefont {I.}~\bibnamefont
  {Bloch}},\ }\href {https://doi.org/10.1038/nature07126} {\bibfield  {journal}
  {\bibinfo  {journal} {Nature}\ }\textbf {\bibinfo {volume} {453}},\ \bibinfo
  {pages} {1016 EP } (\bibinfo {year} {2008})}\BibitemShut {NoStop}%
\bibitem [{\citenamefont {Laflorencie}(2016)}]{2016LaflorenciePhysRep}%
  \BibitemOpen
  \bibfield  {author} {\bibinfo {author} {\bibfnamefont {N.}~\bibnamefont
  {Laflorencie}},\ }\href {\doibase
  https://doi.org/10.1016/j.physrep.2016.06.008} {\bibfield  {journal}
  {\bibinfo  {journal} {Physics Reports}\ }\textbf {\bibinfo {volume} {646}},\
  \bibinfo {pages} {1 } (\bibinfo {year} {2016})},\ \bibinfo {note} {quantum
  entanglement in condensed matter systems}\BibitemShut {NoStop}%
\bibitem [{\citenamefont {Sachdev}(2011)}]{Sachdev2011QPT}%
  \BibitemOpen
  \bibfield  {author} {\bibinfo {author} {\bibfnamefont {S.}~\bibnamefont
  {Sachdev}},\ }\href@noop {} {\emph {\bibinfo {title} {Quantum Phase
  Transitions}}}\ (\bibinfo  {publisher} {Cambridge University Press},\
  \bibinfo {address} {Cambridge},\ \bibinfo {year} {2011})\BibitemShut
  {NoStop}%
\bibitem [{\citenamefont {Osterloh}\ \emph {et~al.}(2002)\citenamefont
  {Osterloh}, \citenamefont {Amico}, \citenamefont {Falci},\ and\ \citenamefont
  {Fazio}}]{osterloh2002scaling}%
  \BibitemOpen
  \bibfield  {author} {\bibinfo {author} {\bibfnamefont {A.}~\bibnamefont
  {Osterloh}}, \bibinfo {author} {\bibfnamefont {L.}~\bibnamefont {Amico}},
  \bibinfo {author} {\bibfnamefont {G.}~\bibnamefont {Falci}}, \ and\ \bibinfo
  {author} {\bibfnamefont {R.}~\bibnamefont {Fazio}},\ }\href {\doibase
  10.1038/416608a} {\bibfield  {journal} {\bibinfo  {journal} {Nature}\
  }\textbf {\bibinfo {volume} {416}},\ \bibinfo {pages} {608} (\bibinfo {year}
  {2002})}\BibitemShut {NoStop}%
\bibitem [{\citenamefont {Vidal}\ \emph {et~al.}(2003)\citenamefont {Vidal},
  \citenamefont {Latorre}, \citenamefont {Rico},\ and\ \citenamefont
  {Kitaev}}]{vidal2003entanglement}%
  \BibitemOpen
  \bibfield  {author} {\bibinfo {author} {\bibfnamefont {G.}~\bibnamefont
  {Vidal}}, \bibinfo {author} {\bibfnamefont {J.~I.}\ \bibnamefont {Latorre}},
  \bibinfo {author} {\bibfnamefont {E.}~\bibnamefont {Rico}}, \ and\ \bibinfo
  {author} {\bibfnamefont {A.}~\bibnamefont {Kitaev}},\ }\href {\doibase
  10.1103/PhysRevLett.90.227902} {\bibfield  {journal} {\bibinfo  {journal}
  {Phys. Rev. Lett.}\ }\textbf {\bibinfo {volume} {90}},\ \bibinfo {pages}
  {227902} (\bibinfo {year} {2003})}\BibitemShut {NoStop}%
\bibitem [{\citenamefont {Miranowicz}\ \emph {et~al.}(2002)\citenamefont
  {Miranowicz}, \citenamefont {\"Ozdemir}, \citenamefont {Liu}, \citenamefont
  {Koashi}, \citenamefont {Imoto},\ and\ \citenamefont
  {Hirayama}}]{miranowicz2002generation}%
  \BibitemOpen
  \bibfield  {author} {\bibinfo {author} {\bibfnamefont {A.}~\bibnamefont
  {Miranowicz}}, \bibinfo {author} {\bibfnamefont {i.~m. c.~K.}\ \bibnamefont
  {\"Ozdemir}}, \bibinfo {author} {\bibfnamefont {Y.-x.}\ \bibnamefont {Liu}},
  \bibinfo {author} {\bibfnamefont {M.}~\bibnamefont {Koashi}}, \bibinfo
  {author} {\bibfnamefont {N.}~\bibnamefont {Imoto}}, \ and\ \bibinfo {author}
  {\bibfnamefont {Y.}~\bibnamefont {Hirayama}},\ }\href {\doibase
  10.1103/PhysRevA.65.062321} {\bibfield  {journal} {\bibinfo  {journal} {Phys.
  Rev. A}\ }\textbf {\bibinfo {volume} {65}},\ \bibinfo {pages} {062321}
  (\bibinfo {year} {2002})}\BibitemShut {NoStop}%
\bibitem [{\citenamefont {Wu}\ \emph {et~al.}(2004)\citenamefont {Wu},
  \citenamefont {Sarandy},\ and\ \citenamefont {Lidar}}]{wu2004quantum}%
  \BibitemOpen
  \bibfield  {author} {\bibinfo {author} {\bibfnamefont {L.-A.}\ \bibnamefont
  {Wu}}, \bibinfo {author} {\bibfnamefont {M.~S.}\ \bibnamefont {Sarandy}}, \
  and\ \bibinfo {author} {\bibfnamefont {D.~A.}\ \bibnamefont {Lidar}},\ }\href
  {\doibase 10.1103/PhysRevLett.93.250404} {\bibfield  {journal} {\bibinfo
  {journal} {Phys. Rev. Lett.}\ }\textbf {\bibinfo {volume} {93}},\ \bibinfo
  {pages} {250404} (\bibinfo {year} {2004})}\BibitemShut {NoStop}%
\bibitem [{\citenamefont {Liu}\ \emph {et~al.}(2013)\citenamefont {Liu},
  \citenamefont {Ma},\ and\ \citenamefont {Wang}}]{liu2013quantum}%
  \BibitemOpen
  \bibfield  {author} {\bibinfo {author} {\bibfnamefont {W.-F.}\ \bibnamefont
  {Liu}}, \bibinfo {author} {\bibfnamefont {J.}~\bibnamefont {Ma}}, \ and\
  \bibinfo {author} {\bibfnamefont {X.}~\bibnamefont {Wang}},\ }\href {\doibase
  10.1088/1751-8113/46/4/045302} {\bibfield  {journal} {\bibinfo  {journal}
  {Journal of Physics A: Mathematical and Theoretical}\ }\textbf {\bibinfo
  {volume} {46}},\ \bibinfo {pages} {045302} (\bibinfo {year}
  {2013})}\BibitemShut {NoStop}%
\bibitem [{\citenamefont {Ma}\ and\ \citenamefont {Wang}(2009)}]{ma2009fisher}%
  \BibitemOpen
  \bibfield  {author} {\bibinfo {author} {\bibfnamefont {J.}~\bibnamefont
  {Ma}}\ and\ \bibinfo {author} {\bibfnamefont {X.}~\bibnamefont {Wang}},\
  }\href {\doibase 10.1103/PhysRevA.80.012318} {\bibfield  {journal} {\bibinfo
  {journal} {Phys. Rev. A}\ }\textbf {\bibinfo {volume} {80}},\ \bibinfo
  {pages} {012318} (\bibinfo {year} {2009})}\BibitemShut {NoStop}%
\bibitem [{\citenamefont {Luo}\ \emph {et~al.}(2017)\citenamefont {Luo},
  \citenamefont {Zou}, \citenamefont {Wu}, \citenamefont {Liu}, \citenamefont
  {Han}, \citenamefont {Tey},\ and\ \citenamefont
  {You}}]{luo2017deterministic}%
  \BibitemOpen
  \bibfield  {author} {\bibinfo {author} {\bibfnamefont {X.-Y.}\ \bibnamefont
  {Luo}}, \bibinfo {author} {\bibfnamefont {Y.-Q.}\ \bibnamefont {Zou}},
  \bibinfo {author} {\bibfnamefont {L.-N.}\ \bibnamefont {Wu}}, \bibinfo
  {author} {\bibfnamefont {Q.}~\bibnamefont {Liu}}, \bibinfo {author}
  {\bibfnamefont {M.-F.}\ \bibnamefont {Han}}, \bibinfo {author} {\bibfnamefont
  {M.~K.}\ \bibnamefont {Tey}}, \ and\ \bibinfo {author} {\bibfnamefont
  {L.}~\bibnamefont {You}},\ }\href {\doibase 10.1126/science.aag1106}
  {\bibfield  {journal} {\bibinfo  {journal} {Science}\ }\textbf {\bibinfo
  {volume} {355}},\ \bibinfo {pages} {620} (\bibinfo {year}
  {2017})}\BibitemShut {NoStop}%
\bibitem [{\citenamefont {Daley}\ \emph {et~al.}(2012)\citenamefont {Daley},
  \citenamefont {Pichler}, \citenamefont {Schachenmayer},\ and\ \citenamefont
  {Zoller}}]{daley2012measuring}%
  \BibitemOpen
  \bibfield  {author} {\bibinfo {author} {\bibfnamefont {A.~J.}\ \bibnamefont
  {Daley}}, \bibinfo {author} {\bibfnamefont {H.}~\bibnamefont {Pichler}},
  \bibinfo {author} {\bibfnamefont {J.}~\bibnamefont {Schachenmayer}}, \ and\
  \bibinfo {author} {\bibfnamefont {P.}~\bibnamefont {Zoller}},\ }\href
  {\doibase 10.1103/PhysRevLett.109.020505} {\bibfield  {journal} {\bibinfo
  {journal} {Phys. Rev. Lett.}\ }\textbf {\bibinfo {volume} {109}},\ \bibinfo
  {pages} {020505} (\bibinfo {year} {2012})}\BibitemShut {NoStop}%
\bibitem [{\citenamefont {Lanting}\ \emph {et~al.}(2014)\citenamefont
  {Lanting}, \citenamefont {Przybysz}, \citenamefont {Smirnov}, \citenamefont
  {Spedalieri}, \citenamefont {Amin}, \citenamefont {Berkley}, \citenamefont
  {Harris}, \citenamefont {Altomare}, \citenamefont {Boixo}, \citenamefont
  {Bunyk}, \citenamefont {Dickson}, \citenamefont {Enderud}, \citenamefont
  {Hilton}, \citenamefont {Hoskinson}, \citenamefont {Johnson}, \citenamefont
  {Ladizinsky}, \citenamefont {Ladizinsky}, \citenamefont {Neufeld},
  \citenamefont {Oh}, \citenamefont {Perminov}, \citenamefont {Rich},
  \citenamefont {Thom}, \citenamefont {Tolkacheva}, \citenamefont {Uchaikin},
  \citenamefont {Wilson},\ and\ \citenamefont
  {Rose}}]{lanting2014entanglement}%
  \BibitemOpen
  \bibfield  {author} {\bibinfo {author} {\bibfnamefont {T.}~\bibnamefont
  {Lanting}}, \bibinfo {author} {\bibfnamefont {A.~J.}\ \bibnamefont
  {Przybysz}}, \bibinfo {author} {\bibfnamefont {A.~Y.}\ \bibnamefont
  {Smirnov}}, \bibinfo {author} {\bibfnamefont {F.~M.}\ \bibnamefont
  {Spedalieri}}, \bibinfo {author} {\bibfnamefont {M.~H.}\ \bibnamefont
  {Amin}}, \bibinfo {author} {\bibfnamefont {A.~J.}\ \bibnamefont {Berkley}},
  \bibinfo {author} {\bibfnamefont {R.}~\bibnamefont {Harris}}, \bibinfo
  {author} {\bibfnamefont {F.}~\bibnamefont {Altomare}}, \bibinfo {author}
  {\bibfnamefont {S.}~\bibnamefont {Boixo}}, \bibinfo {author} {\bibfnamefont
  {P.}~\bibnamefont {Bunyk}}, \bibinfo {author} {\bibfnamefont
  {N.}~\bibnamefont {Dickson}}, \bibinfo {author} {\bibfnamefont
  {C.}~\bibnamefont {Enderud}}, \bibinfo {author} {\bibfnamefont {J.~P.}\
  \bibnamefont {Hilton}}, \bibinfo {author} {\bibfnamefont {E.}~\bibnamefont
  {Hoskinson}}, \bibinfo {author} {\bibfnamefont {M.~W.}\ \bibnamefont
  {Johnson}}, \bibinfo {author} {\bibfnamefont {E.}~\bibnamefont {Ladizinsky}},
  \bibinfo {author} {\bibfnamefont {N.}~\bibnamefont {Ladizinsky}}, \bibinfo
  {author} {\bibfnamefont {R.}~\bibnamefont {Neufeld}}, \bibinfo {author}
  {\bibfnamefont {T.}~\bibnamefont {Oh}}, \bibinfo {author} {\bibfnamefont
  {I.}~\bibnamefont {Perminov}}, \bibinfo {author} {\bibfnamefont
  {C.}~\bibnamefont {Rich}}, \bibinfo {author} {\bibfnamefont {M.~C.}\
  \bibnamefont {Thom}}, \bibinfo {author} {\bibfnamefont {E.}~\bibnamefont
  {Tolkacheva}}, \bibinfo {author} {\bibfnamefont {S.}~\bibnamefont
  {Uchaikin}}, \bibinfo {author} {\bibfnamefont {A.~B.}\ \bibnamefont
  {Wilson}}, \ and\ \bibinfo {author} {\bibfnamefont {G.}~\bibnamefont
  {Rose}},\ }\href {\doibase 10.1103/PhysRevX.4.021041} {\bibfield  {journal}
  {\bibinfo  {journal} {Phys. Rev. X}\ }\textbf {\bibinfo {volume} {4}},\
  \bibinfo {pages} {021041} (\bibinfo {year} {2014})}\BibitemShut {NoStop}%
\bibitem [{\citenamefont {Islam}\ \emph {et~al.}(2015)\citenamefont {Islam},
  \citenamefont {Ma}, \citenamefont {Preiss}, \citenamefont {Eric~Tai},
  \citenamefont {Lukin}, \citenamefont {Rispoli},\ and\ \citenamefont
  {Greiner}}]{islam2015measuring}%
  \BibitemOpen
  \bibfield  {author} {\bibinfo {author} {\bibfnamefont {R.}~\bibnamefont
  {Islam}}, \bibinfo {author} {\bibfnamefont {R.}~\bibnamefont {Ma}}, \bibinfo
  {author} {\bibfnamefont {P.~M.}\ \bibnamefont {Preiss}}, \bibinfo {author}
  {\bibfnamefont {M.}~\bibnamefont {Eric~Tai}}, \bibinfo {author}
  {\bibfnamefont {A.}~\bibnamefont {Lukin}}, \bibinfo {author} {\bibfnamefont
  {M.}~\bibnamefont {Rispoli}}, \ and\ \bibinfo {author} {\bibfnamefont
  {M.}~\bibnamefont {Greiner}},\ }\href {https://doi.org/10.1038/nature15750}
  {\bibfield  {journal} {\bibinfo  {journal} {Nature}\ }\textbf {\bibinfo
  {volume} {528}},\ \bibinfo {pages} {77 EP } (\bibinfo {year}
  {2015})}\BibitemShut {NoStop}%
\bibitem [{\citenamefont {Fukuhara}\ \emph {et~al.}(2015)\citenamefont
  {Fukuhara}, \citenamefont {Hild}, \citenamefont {Zeiher}, \citenamefont
  {Schau\ss{}}, \citenamefont {Bloch}, \citenamefont {Endres},\ and\
  \citenamefont {Gross}}]{fukuhara2015spatially}%
  \BibitemOpen
  \bibfield  {author} {\bibinfo {author} {\bibfnamefont {T.}~\bibnamefont
  {Fukuhara}}, \bibinfo {author} {\bibfnamefont {S.}~\bibnamefont {Hild}},
  \bibinfo {author} {\bibfnamefont {J.}~\bibnamefont {Zeiher}}, \bibinfo
  {author} {\bibfnamefont {P.}~\bibnamefont {Schau\ss{}}}, \bibinfo {author}
  {\bibfnamefont {I.}~\bibnamefont {Bloch}}, \bibinfo {author} {\bibfnamefont
  {M.}~\bibnamefont {Endres}}, \ and\ \bibinfo {author} {\bibfnamefont
  {C.}~\bibnamefont {Gross}},\ }\href {\doibase 10.1103/PhysRevLett.115.035302}
  {\bibfield  {journal} {\bibinfo  {journal} {Phys. Rev. Lett.}\ }\textbf
  {\bibinfo {volume} {115}},\ \bibinfo {pages} {035302} (\bibinfo {year}
  {2015})}\BibitemShut {NoStop}%
\bibitem [{\citenamefont {Wie{\'{s}}niak}\ \emph {et~al.}(2005)\citenamefont
  {Wie{\'{s}}niak}, \citenamefont {Vedral},\ and\ \citenamefont
  {Brukner}}]{2005WiesniakNJP}%
  \BibitemOpen
  \bibfield  {author} {\bibinfo {author} {\bibfnamefont {M.}~\bibnamefont
  {Wie{\'{s}}niak}}, \bibinfo {author} {\bibfnamefont {V.}~\bibnamefont
  {Vedral}}, \ and\ \bibinfo {author} {\bibfnamefont {{\v{C}}.}~\bibnamefont
  {Brukner}},\ }\href {\doibase 10.1088/1367-2630/7/1/258} {\bibfield
  {journal} {\bibinfo  {journal} {New Journal of Physics}\ }\textbf {\bibinfo
  {volume} {7}},\ \bibinfo {pages} {258} (\bibinfo {year} {2005})}\BibitemShut
  {NoStop}%
\bibitem [{\citenamefont {Hauke}\ \emph {et~al.}(2016)\citenamefont {Hauke},
  \citenamefont {Heyl}, \citenamefont {Tagliacozzo},\ and\ \citenamefont
  {Zoller}}]{hauke2016measuring}%
  \BibitemOpen
  \bibfield  {author} {\bibinfo {author} {\bibfnamefont {P.}~\bibnamefont
  {Hauke}}, \bibinfo {author} {\bibfnamefont {M.}~\bibnamefont {Heyl}},
  \bibinfo {author} {\bibfnamefont {L.}~\bibnamefont {Tagliacozzo}}, \ and\
  \bibinfo {author} {\bibfnamefont {P.}~\bibnamefont {Zoller}},\ }\href@noop {}
  {\bibfield  {journal} {\bibinfo  {journal} {Nature Physics}\ }\textbf
  {\bibinfo {volume} {12}},\ \bibinfo {pages} {778} (\bibinfo {year}
  {2016})}\BibitemShut {NoStop}%
\bibitem [{\citenamefont {Fisher}(1974)}]{1974FisherRMP}%
  \BibitemOpen
  \bibfield  {author} {\bibinfo {author} {\bibfnamefont {M.~E.}\ \bibnamefont
  {Fisher}},\ }\href {\doibase 10.1103/RevModPhys.46.597} {\bibfield  {journal}
  {\bibinfo  {journal} {Rev. Mod. Phys.}\ }\textbf {\bibinfo {volume} {46}},\
  \bibinfo {pages} {597} (\bibinfo {year} {1974})}\BibitemShut {NoStop}%
\bibitem [{\citenamefont {P\ifmmode~\mbox{\k{e}}\else \k{e}\fi{}cak}\ and\
  \citenamefont {Sowi\ifmmode~\acute{n}\else
  \'{n}\fi{}ski}(2016)}]{2016PecakPRA}%
  \BibitemOpen
  \bibfield  {author} {\bibinfo {author} {\bibfnamefont {D.}~\bibnamefont
  {P\ifmmode~\mbox{\k{e}}\else \k{e}\fi{}cak}}\ and\ \bibinfo {author}
  {\bibfnamefont {T.}~\bibnamefont {Sowi\ifmmode~\acute{n}\else
  \'{n}\fi{}ski}},\ }\href {\doibase 10.1103/PhysRevA.94.042118} {\bibfield
  {journal} {\bibinfo  {journal} {Phys. Rev. A}\ }\textbf {\bibinfo {volume}
  {94}},\ \bibinfo {pages} {042118} (\bibinfo {year} {2016})}\BibitemShut
  {NoStop}%
\bibitem [{\citenamefont {Serwane}\ \emph {et~al.}(2011)\citenamefont
  {Serwane}, \citenamefont {Z{\"u}rn}, \citenamefont {Lompe}, \citenamefont
  {Ottenstein}, \citenamefont {Wenz},\ and\ \citenamefont
  {Jochim}}]{2011SerwaneScience}%
  \BibitemOpen
  \bibfield  {author} {\bibinfo {author} {\bibfnamefont {F.}~\bibnamefont
  {Serwane}}, \bibinfo {author} {\bibfnamefont {G.}~\bibnamefont {Z{\"u}rn}},
  \bibinfo {author} {\bibfnamefont {T.}~\bibnamefont {Lompe}}, \bibinfo
  {author} {\bibfnamefont {T.~B.}\ \bibnamefont {Ottenstein}}, \bibinfo
  {author} {\bibfnamefont {A.~N.}\ \bibnamefont {Wenz}}, \ and\ \bibinfo
  {author} {\bibfnamefont {S.}~\bibnamefont {Jochim}},\ }\href {\doibase
  10.1126/science.1201351} {\bibfield  {journal} {\bibinfo  {journal}
  {Science}\ }\textbf {\bibinfo {volume} {332}},\ \bibinfo {pages} {336}
  (\bibinfo {year} {2011})}\BibitemShut {NoStop}%
\bibitem [{\citenamefont {Wenz}\ \emph {et~al.}(2013)\citenamefont {Wenz},
  \citenamefont {Z{\"u}rn}, \citenamefont {Murmann}, \citenamefont {Brouzos},
  \citenamefont {Lompe},\ and\ \citenamefont {Jochim}}]{2013WenzScience}%
  \BibitemOpen
  \bibfield  {author} {\bibinfo {author} {\bibfnamefont {A.~N.}\ \bibnamefont
  {Wenz}}, \bibinfo {author} {\bibfnamefont {G.}~\bibnamefont {Z{\"u}rn}},
  \bibinfo {author} {\bibfnamefont {S.}~\bibnamefont {Murmann}}, \bibinfo
  {author} {\bibfnamefont {I.}~\bibnamefont {Brouzos}}, \bibinfo {author}
  {\bibfnamefont {T.}~\bibnamefont {Lompe}}, \ and\ \bibinfo {author}
  {\bibfnamefont {S.}~\bibnamefont {Jochim}},\ }\href {\doibase
  10.1126/science.1240516} {\bibfield  {journal} {\bibinfo  {journal}
  {Science}\ }\textbf {\bibinfo {volume} {342}},\ \bibinfo {pages} {457}
  (\bibinfo {year} {2013})}\BibitemShut {NoStop}%
\bibitem [{\citenamefont {Z\"urn}\ \emph {et~al.}(2013)\citenamefont {Z\"urn},
  \citenamefont {Wenz}, \citenamefont {Murmann}, \citenamefont {Bergschneider},
  \citenamefont {Lompe},\ and\ \citenamefont {Jochim}}]{2013ZurnPRL}%
  \BibitemOpen
  \bibfield  {author} {\bibinfo {author} {\bibfnamefont {G.}~\bibnamefont
  {Z\"urn}}, \bibinfo {author} {\bibfnamefont {A.~N.}\ \bibnamefont {Wenz}},
  \bibinfo {author} {\bibfnamefont {S.}~\bibnamefont {Murmann}}, \bibinfo
  {author} {\bibfnamefont {A.}~\bibnamefont {Bergschneider}}, \bibinfo {author}
  {\bibfnamefont {T.}~\bibnamefont {Lompe}}, \ and\ \bibinfo {author}
  {\bibfnamefont {S.}~\bibnamefont {Jochim}},\ }\href {\doibase
  10.1103/PhysRevLett.111.175302} {\bibfield  {journal} {\bibinfo  {journal}
  {Phys. Rev. Lett.}\ }\textbf {\bibinfo {volume} {111}},\ \bibinfo {pages}
  {175302} (\bibinfo {year} {2013})}\BibitemShut {NoStop}%
\bibitem [{\citenamefont {Tiecke}\ \emph {et~al.}(2010)\citenamefont {Tiecke},
  \citenamefont {Goosen}, \citenamefont {Ludewig}, \citenamefont {Gensemer},
  \citenamefont {Kraft}, \citenamefont {Kokkelmans},\ and\ \citenamefont
  {Walraven}}]{2010TieckePRL}%
  \BibitemOpen
  \bibfield  {author} {\bibinfo {author} {\bibfnamefont {T.~G.}\ \bibnamefont
  {Tiecke}}, \bibinfo {author} {\bibfnamefont {M.~R.}\ \bibnamefont {Goosen}},
  \bibinfo {author} {\bibfnamefont {A.}~\bibnamefont {Ludewig}}, \bibinfo
  {author} {\bibfnamefont {S.~D.}\ \bibnamefont {Gensemer}}, \bibinfo {author}
  {\bibfnamefont {S.}~\bibnamefont {Kraft}}, \bibinfo {author} {\bibfnamefont
  {S.~J. J. M.~F.}\ \bibnamefont {Kokkelmans}}, \ and\ \bibinfo {author}
  {\bibfnamefont {J.~T.~M.}\ \bibnamefont {Walraven}},\ }\href {\doibase
  10.1103/PhysRevLett.104.053202} {\bibfield  {journal} {\bibinfo  {journal}
  {Phys. Rev. Lett.}\ }\textbf {\bibinfo {volume} {104}},\ \bibinfo {pages}
  {053202} (\bibinfo {year} {2010})}\BibitemShut {NoStop}%
\bibitem [{\citenamefont {Naik}\ \emph {et~al.}(2011)\citenamefont {Naik},
  \citenamefont {Trenkwalder}, \citenamefont {Kohstall}, \citenamefont
  {Spiegelhalder}, \citenamefont {Zaccanti}, \citenamefont {Hendl},
  \citenamefont {Schreck}, \citenamefont {Grimm}, \citenamefont {Hanna},\ and\
  \citenamefont {Julienne}}]{2011NaikEPJD}%
  \BibitemOpen
  \bibfield  {author} {\bibinfo {author} {\bibfnamefont {D.}~\bibnamefont
  {Naik}}, \bibinfo {author} {\bibfnamefont {A.}~\bibnamefont {Trenkwalder}},
  \bibinfo {author} {\bibfnamefont {C.}~\bibnamefont {Kohstall}}, \bibinfo
  {author} {\bibfnamefont {F.~M.}\ \bibnamefont {Spiegelhalder}}, \bibinfo
  {author} {\bibfnamefont {M.}~\bibnamefont {Zaccanti}}, \bibinfo {author}
  {\bibfnamefont {G.}~\bibnamefont {Hendl}}, \bibinfo {author} {\bibfnamefont
  {F.}~\bibnamefont {Schreck}}, \bibinfo {author} {\bibfnamefont
  {R.}~\bibnamefont {Grimm}}, \bibinfo {author} {\bibfnamefont {T.~M.}\
  \bibnamefont {Hanna}}, \ and\ \bibinfo {author} {\bibfnamefont {P.~S.}\
  \bibnamefont {Julienne}},\ }\href {\doibase 10.1140/epjd/e2010-10591-2}
  {\bibfield  {journal} {\bibinfo  {journal} {The European Physical Journal D}\
  }\textbf {\bibinfo {volume} {65}},\ \bibinfo {pages} {55} (\bibinfo {year}
  {2011})}\BibitemShut {NoStop}%
\bibitem [{\citenamefont {Cetina}\ \emph {et~al.}(2015)\citenamefont {Cetina},
  \citenamefont {Jag}, \citenamefont {Lous}, \citenamefont {Walraven},
  \citenamefont {Grimm}, \citenamefont {Christensen},\ and\ \citenamefont
  {Bruun}}]{2015CetinaPRL}%
  \BibitemOpen
  \bibfield  {author} {\bibinfo {author} {\bibfnamefont {M.}~\bibnamefont
  {Cetina}}, \bibinfo {author} {\bibfnamefont {M.}~\bibnamefont {Jag}},
  \bibinfo {author} {\bibfnamefont {R.~S.}\ \bibnamefont {Lous}}, \bibinfo
  {author} {\bibfnamefont {J.~T.~M.}\ \bibnamefont {Walraven}}, \bibinfo
  {author} {\bibfnamefont {R.}~\bibnamefont {Grimm}}, \bibinfo {author}
  {\bibfnamefont {R.~S.}\ \bibnamefont {Christensen}}, \ and\ \bibinfo {author}
  {\bibfnamefont {G.~M.}\ \bibnamefont {Bruun}},\ }\href {\doibase
  10.1103/PhysRevLett.115.135302} {\bibfield  {journal} {\bibinfo  {journal}
  {Phys. Rev. Lett.}\ }\textbf {\bibinfo {volume} {115}},\ \bibinfo {pages}
  {135302} (\bibinfo {year} {2015})}\BibitemShut {NoStop}%
\bibitem [{\citenamefont {Cetina}\ \emph {et~al.}(2016)\citenamefont {Cetina},
  \citenamefont {Jag}, \citenamefont {Lous}, \citenamefont {Fritsche},
  \citenamefont {Walraven}, \citenamefont {Grimm}, \citenamefont {Levinsen},
  \citenamefont {Parish}, \citenamefont {Schmidt}, \citenamefont {Knap},\ and\
  \citenamefont {Demler}}]{cetina2016ultrafast}%
  \BibitemOpen
  \bibfield  {author} {\bibinfo {author} {\bibfnamefont {M.}~\bibnamefont
  {Cetina}}, \bibinfo {author} {\bibfnamefont {M.}~\bibnamefont {Jag}},
  \bibinfo {author} {\bibfnamefont {R.~S.}\ \bibnamefont {Lous}}, \bibinfo
  {author} {\bibfnamefont {I.}~\bibnamefont {Fritsche}}, \bibinfo {author}
  {\bibfnamefont {J.~T.~M.}\ \bibnamefont {Walraven}}, \bibinfo {author}
  {\bibfnamefont {R.}~\bibnamefont {Grimm}}, \bibinfo {author} {\bibfnamefont
  {J.}~\bibnamefont {Levinsen}}, \bibinfo {author} {\bibfnamefont {M.~M.}\
  \bibnamefont {Parish}}, \bibinfo {author} {\bibfnamefont {R.}~\bibnamefont
  {Schmidt}}, \bibinfo {author} {\bibfnamefont {M.}~\bibnamefont {Knap}}, \
  and\ \bibinfo {author} {\bibfnamefont {E.}~\bibnamefont {Demler}},\ }\href
  {\doibase 10.1126/science.aaf5134} {\bibfield  {journal} {\bibinfo  {journal}
  {Science}\ }\textbf {\bibinfo {volume} {354}},\ \bibinfo {pages} {96}
  (\bibinfo {year} {2016})}\BibitemShut {NoStop}%
\bibitem [{\citenamefont {Ravensbergen}\ \emph {et~al.}(2018)\citenamefont
  {Ravensbergen}, \citenamefont {Corre}, \citenamefont {Soave}, \citenamefont
  {Kreyer}, \citenamefont {Kirilov},\ and\ \citenamefont
  {Grimm}}]{Grimm2018DyK}%
  \BibitemOpen
  \bibfield  {author} {\bibinfo {author} {\bibfnamefont {C.}~\bibnamefont
  {Ravensbergen}}, \bibinfo {author} {\bibfnamefont {V.}~\bibnamefont {Corre}},
  \bibinfo {author} {\bibfnamefont {E.}~\bibnamefont {Soave}}, \bibinfo
  {author} {\bibfnamefont {M.}~\bibnamefont {Kreyer}}, \bibinfo {author}
  {\bibfnamefont {E.}~\bibnamefont {Kirilov}}, \ and\ \bibinfo {author}
  {\bibfnamefont {R.}~\bibnamefont {Grimm}},\ }\href {\doibase
  10.1103/PhysRevA.98.063624} {\bibfield  {journal} {\bibinfo  {journal} {Phys.
  Rev. A}\ }\textbf {\bibinfo {volume} {98}},\ \bibinfo {pages} {063624}
  (\bibinfo {year} {2018})}\BibitemShut {NoStop}%
\bibitem [{\citenamefont {Sowi{\'{n}}ski}\ and\ \citenamefont
  {Garc{\'{\i}}a-March}(2019)}]{2019SowinskiRPP}%
  \BibitemOpen
  \bibfield  {author} {\bibinfo {author} {\bibfnamefont {T.}~\bibnamefont
  {Sowi{\'{n}}ski}}\ and\ \bibinfo {author} {\bibfnamefont {M.~{\'{A}}.}\
  \bibnamefont {Garc{\'{\i}}a-March}},\ }\href {\doibase
  10.1088/1361-6633/ab3a80} {\bibfield  {journal} {\bibinfo  {journal} {Reports
  on Progress in Physics}\ }\textbf {\bibinfo {volume} {82}},\ \bibinfo {pages}
  {104401} (\bibinfo {year} {2019})}\BibitemShut {NoStop}%
\bibitem [{\citenamefont {Mistakidis}\ \emph {et~al.}(2019)\citenamefont
  {Mistakidis}, \citenamefont {Katsimiga}, \citenamefont {Koutentakis},\ and\
  \citenamefont {Schmelcher}}]{2019MistakidisNJP}%
  \BibitemOpen
  \bibfield  {author} {\bibinfo {author} {\bibfnamefont {S.~I.}\ \bibnamefont
  {Mistakidis}}, \bibinfo {author} {\bibfnamefont {G.~C.}\ \bibnamefont
  {Katsimiga}}, \bibinfo {author} {\bibfnamefont {G.~M.}\ \bibnamefont
  {Koutentakis}}, \ and\ \bibinfo {author} {\bibfnamefont {P.}~\bibnamefont
  {Schmelcher}},\ }\href {\doibase 10.1088/1367-2630/ab1045} {\bibfield
  {journal} {\bibinfo  {journal} {New Journal of Physics}\ }\textbf {\bibinfo
  {volume} {21}},\ \bibinfo {pages} {043032} (\bibinfo {year}
  {2019})}\BibitemShut {NoStop}%
\bibitem [{Sup()}]{Supplement}%
  \BibitemOpen
  \href@noop {} {}\bibinfo {note} {See supplementary material for results
  obtained for other system parameters.}\BibitemShut {Stop}%
\bibitem [{\citenamefont {Haugset}\ and\ \citenamefont
  {Haugerud}(1998)}]{1998HaugsetPRA}%
  \BibitemOpen
  \bibfield  {author} {\bibinfo {author} {\bibfnamefont {T.}~\bibnamefont
  {Haugset}}\ and\ \bibinfo {author} {\bibfnamefont {H.}~\bibnamefont
  {Haugerud}},\ }\href {\doibase 10.1103/PhysRevA.57.3809} {\bibfield
  {journal} {\bibinfo  {journal} {Phys. Rev. A}\ }\textbf {\bibinfo {volume}
  {57}},\ \bibinfo {pages} {3809} (\bibinfo {year} {1998})}\BibitemShut
  {NoStop}%
\bibitem [{\citenamefont {Chrostowski}\ and\ \citenamefont
  {Sowi\'nski}(2019)}]{2019ChrostowskiAPPA}%
  \BibitemOpen
  \bibfield  {author} {\bibinfo {author} {\bibfnamefont {A.}~\bibnamefont
  {Chrostowski}}\ and\ \bibinfo {author} {\bibfnamefont {T.}~\bibnamefont
  {Sowi\'nski}},\ }\href {\doibase 10.12693/APhysPolA.136.566} {\bibfield
  {journal} {\bibinfo  {journal} {Acta Phys. Polon. A}\ }\textbf {\bibinfo
  {volume} {136}},\ \bibinfo {pages} {566} (\bibinfo {year}
  {2019})}\BibitemShut {NoStop}%
\bibitem [{\citenamefont {Lehoucq}\ \emph {et~al.}(1998)\citenamefont
  {Lehoucq}, \citenamefont {Sorensen},\ and\ \citenamefont
  {Yang}}]{ArnoldiBook}%
  \BibitemOpen
  \bibfield  {author} {\bibinfo {author} {\bibfnamefont {R.~B.}\ \bibnamefont
  {Lehoucq}}, \bibinfo {author} {\bibfnamefont {D.~C.}\ \bibnamefont
  {Sorensen}}, \ and\ \bibinfo {author} {\bibfnamefont {C.}~\bibnamefont
  {Yang}},\ }\href@noop {} {\emph {\bibinfo {title} {Arpack Users Guide:
  Solution of Large-Scale Eigenvalue Problems With Implicityly Restorted
  Arnoldi Methods}}}\ (\bibinfo  {publisher} {Society for Industrial \& Applied
  Mathematics},\ \bibinfo {address} {Philadelphia},\ \bibinfo {year}
  {1998})\BibitemShut {NoStop}%
\bibitem [{\citenamefont {Loft}\ \emph {et~al.}(2015)\citenamefont {Loft},
  \citenamefont {Dehkharghani}, \citenamefont {Mehta}, \citenamefont
  {Volosniev},\ and\ \citenamefont {Zinner}}]{2015LoftEPJD}%
  \BibitemOpen
  \bibfield  {author} {\bibinfo {author} {\bibfnamefont {N.~J.~S.}\
  \bibnamefont {Loft}}, \bibinfo {author} {\bibfnamefont {A.~S.}\ \bibnamefont
  {Dehkharghani}}, \bibinfo {author} {\bibfnamefont {N.~P.}\ \bibnamefont
  {Mehta}}, \bibinfo {author} {\bibfnamefont {A.~G.}\ \bibnamefont
  {Volosniev}}, \ and\ \bibinfo {author} {\bibfnamefont {N.~T.}\ \bibnamefont
  {Zinner}},\ }\href {\doibase 10.1140/epjd/e2015-50845-9} {\bibfield
  {journal} {\bibinfo  {journal} {The European Physical Journal D}\ }\textbf
  {\bibinfo {volume} {69}},\ \bibinfo {pages} {65} (\bibinfo {year}
  {2015})}\BibitemShut {NoStop}%
\bibitem [{\citenamefont {P{\c e}cak}\ \emph {et~al.}(2016)\citenamefont {P{\c
  e}cak}, \citenamefont {Gajda},\ and\ \citenamefont
  {Sowi{\'n}ski}}]{2016PecakNJP}%
  \BibitemOpen
  \bibfield  {author} {\bibinfo {author} {\bibfnamefont {D.}~\bibnamefont {P{\c
  e}cak}}, \bibinfo {author} {\bibfnamefont {M.}~\bibnamefont {Gajda}}, \ and\
  \bibinfo {author} {\bibfnamefont {T.}~\bibnamefont {Sowi{\'n}ski}},\ }\href
  {http://stacks.iop.org/1367-2630/18/i=1/a=013030} {\bibfield  {journal}
  {\bibinfo  {journal} {New Journal of Physics}\ }\textbf {\bibinfo {volume}
  {18}},\ \bibinfo {pages} {013030} (\bibinfo {year} {2016})}\BibitemShut
  {NoStop}%
\bibitem [{\citenamefont {Dehkharghani}\ \emph {et~al.}(2016)\citenamefont
  {Dehkharghani}, \citenamefont {Volosniev},\ and\ \citenamefont
  {Zinner}}]{2016DehkharghaniJPhysB}%
  \BibitemOpen
  \bibfield  {author} {\bibinfo {author} {\bibfnamefont {A.~S.}\ \bibnamefont
  {Dehkharghani}}, \bibinfo {author} {\bibfnamefont {A.~G.}\ \bibnamefont
  {Volosniev}}, \ and\ \bibinfo {author} {\bibfnamefont {N.~T.}\ \bibnamefont
  {Zinner}},\ }\href {\doibase 10.1088/0953-4075/49/8/085301} {\bibfield
  {journal} {\bibinfo  {journal} {Journal of Physics B: Atomic, Molecular and
  Optical Physics}\ }\textbf {\bibinfo {volume} {49}},\ \bibinfo {pages}
  {085301} (\bibinfo {year} {2016})}\BibitemShut {NoStop}%
\bibitem [{\citenamefont {van Es}\ \emph {et~al.}(2010)\citenamefont {van Es},
  \citenamefont {Wicke}, \citenamefont {van Amerongen}, \citenamefont
  {R{\'{e}}tif}, \citenamefont {Whitlock},\ and\ \citenamefont {van
  Druten}}]{2010vanEsJPhysB}%
  \BibitemOpen
  \bibfield  {author} {\bibinfo {author} {\bibfnamefont {J.~J.~P.}\
  \bibnamefont {van Es}}, \bibinfo {author} {\bibfnamefont {P.}~\bibnamefont
  {Wicke}}, \bibinfo {author} {\bibfnamefont {A.~H.}\ \bibnamefont {van
  Amerongen}}, \bibinfo {author} {\bibfnamefont {C.}~\bibnamefont
  {R{\'{e}}tif}}, \bibinfo {author} {\bibfnamefont {S.}~\bibnamefont
  {Whitlock}}, \ and\ \bibinfo {author} {\bibfnamefont {N.~J.}\ \bibnamefont
  {van Druten}},\ }\href {\doibase 10.1088/0953-4075/43/15/155002} {\bibfield
  {journal} {\bibinfo  {journal} {Journal of Physics B: Atomic, Molecular and
  Optical Physics}\ }\textbf {\bibinfo {volume} {43}},\ \bibinfo {pages}
  {155002} (\bibinfo {year} {2010})}\BibitemShut {NoStop}%
\bibitem [{\citenamefont {Gaunt}\ \emph {et~al.}(2013)\citenamefont {Gaunt},
  \citenamefont {Schmidutz}, \citenamefont {Gotlibovych}, \citenamefont
  {Smith},\ and\ \citenamefont {Hadzibabic}}]{2013GauntPRL}%
  \BibitemOpen
  \bibfield  {author} {\bibinfo {author} {\bibfnamefont {A.~L.}\ \bibnamefont
  {Gaunt}}, \bibinfo {author} {\bibfnamefont {T.~F.}\ \bibnamefont
  {Schmidutz}}, \bibinfo {author} {\bibfnamefont {I.}~\bibnamefont
  {Gotlibovych}}, \bibinfo {author} {\bibfnamefont {R.~P.}\ \bibnamefont
  {Smith}}, \ and\ \bibinfo {author} {\bibfnamefont {Z.}~\bibnamefont
  {Hadzibabic}},\ }\href {\doibase 10.1103/PhysRevLett.110.200406} {\bibfield
  {journal} {\bibinfo  {journal} {Phys. Rev. Lett.}\ }\textbf {\bibinfo
  {volume} {110}},\ \bibinfo {pages} {200406} (\bibinfo {year}
  {2013})}\BibitemShut {NoStop}%
\bibitem [{\citenamefont {Chomaz}\ \emph {et~al.}(2015)\citenamefont {Chomaz},
  \citenamefont {Corman}, \citenamefont {Bienaim{\'e}}, \citenamefont
  {Desbuquois}, \citenamefont {Weitenberg}, \citenamefont {Nascimb{\`e}ne},
  \citenamefont {Beugnon},\ and\ \citenamefont {Dalibard}}]{2015ChomazNatComm}%
  \BibitemOpen
  \bibfield  {author} {\bibinfo {author} {\bibfnamefont {L.}~\bibnamefont
  {Chomaz}}, \bibinfo {author} {\bibfnamefont {L.}~\bibnamefont {Corman}},
  \bibinfo {author} {\bibfnamefont {T.}~\bibnamefont {Bienaim{\'e}}}, \bibinfo
  {author} {\bibfnamefont {R.}~\bibnamefont {Desbuquois}}, \bibinfo {author}
  {\bibfnamefont {C.}~\bibnamefont {Weitenberg}}, \bibinfo {author}
  {\bibfnamefont {S.}~\bibnamefont {Nascimb{\`e}ne}}, \bibinfo {author}
  {\bibfnamefont {J.}~\bibnamefont {Beugnon}}, \ and\ \bibinfo {author}
  {\bibfnamefont {J.}~\bibnamefont {Dalibard}},\ }\href
  {https://doi.org/10.1038/ncomms7162} {\bibfield  {journal} {\bibinfo
  {journal} {Nature Communications}\ }\textbf {\bibinfo {volume} {6}},\
  \bibinfo {pages} {6162 EP } (\bibinfo {year} {2015})}\BibitemShut {NoStop}%
\bibitem [{\citenamefont {Mukherjee}\ \emph {et~al.}(2017)\citenamefont
  {Mukherjee}, \citenamefont {Yan}, \citenamefont {Patel}, \citenamefont
  {Hadzibabic}, \citenamefont {Yefsah}, \citenamefont {Struck},\ and\
  \citenamefont {Zwierlein}}]{2017MukherjeePRL}%
  \BibitemOpen
  \bibfield  {author} {\bibinfo {author} {\bibfnamefont {B.}~\bibnamefont
  {Mukherjee}}, \bibinfo {author} {\bibfnamefont {Z.}~\bibnamefont {Yan}},
  \bibinfo {author} {\bibfnamefont {P.~B.}\ \bibnamefont {Patel}}, \bibinfo
  {author} {\bibfnamefont {Z.}~\bibnamefont {Hadzibabic}}, \bibinfo {author}
  {\bibfnamefont {T.}~\bibnamefont {Yefsah}}, \bibinfo {author} {\bibfnamefont
  {J.}~\bibnamefont {Struck}}, \ and\ \bibinfo {author} {\bibfnamefont {M.~W.}\
  \bibnamefont {Zwierlein}},\ }\href {\doibase 10.1103/PhysRevLett.118.123401}
  {\bibfield  {journal} {\bibinfo  {journal} {Phys. Rev. Lett.}\ }\textbf
  {\bibinfo {volume} {118}},\ \bibinfo {pages} {123401} (\bibinfo {year}
  {2017})}\BibitemShut {NoStop}%
\bibitem [{\citenamefont {Hueck}\ \emph {et~al.}(2018)\citenamefont {Hueck},
  \citenamefont {Luick}, \citenamefont {Sobirey}, \citenamefont {Siegl},
  \citenamefont {Lompe},\ and\ \citenamefont {Moritz}}]{2018HueckPRL}%
  \BibitemOpen
  \bibfield  {author} {\bibinfo {author} {\bibfnamefont {K.}~\bibnamefont
  {Hueck}}, \bibinfo {author} {\bibfnamefont {N.}~\bibnamefont {Luick}},
  \bibinfo {author} {\bibfnamefont {L.}~\bibnamefont {Sobirey}}, \bibinfo
  {author} {\bibfnamefont {J.}~\bibnamefont {Siegl}}, \bibinfo {author}
  {\bibfnamefont {T.}~\bibnamefont {Lompe}}, \ and\ \bibinfo {author}
  {\bibfnamefont {H.}~\bibnamefont {Moritz}},\ }\href {\doibase
  10.1103/PhysRevLett.120.060402} {\bibfield  {journal} {\bibinfo  {journal}
  {Phys. Rev. Lett.}\ }\textbf {\bibinfo {volume} {120}},\ \bibinfo {pages}
  {060402} (\bibinfo {year} {2018})}\BibitemShut {NoStop}%
\bibitem [{\citenamefont {Gu}\ \emph {et~al.}(2008)\citenamefont {Gu},
  \citenamefont {Kwok}, \citenamefont {Ning},\ and\ \citenamefont
  {Lin}}]{2008GuPRA}%
  \BibitemOpen
  \bibfield  {author} {\bibinfo {author} {\bibfnamefont {S.-J.}\ \bibnamefont
  {Gu}}, \bibinfo {author} {\bibfnamefont {H.-M.}\ \bibnamefont {Kwok}},
  \bibinfo {author} {\bibfnamefont {W.-Q.}\ \bibnamefont {Ning}}, \ and\
  \bibinfo {author} {\bibfnamefont {H.-Q.}\ \bibnamefont {Lin}},\ }\href
  {\doibase 10.1103/PhysRevB.77.245109} {\bibfield  {journal} {\bibinfo
  {journal} {Phys. Rev. B}\ }\textbf {\bibinfo {volume} {77}},\ \bibinfo
  {pages} {245109} (\bibinfo {year} {2008})}\BibitemShut {NoStop}%
\bibitem [{\citenamefont {Sowi\ifmmode~\acute{n}\else \'{n}\fi{}ski}\ \emph
  {et~al.}(2015)\citenamefont {Sowi\ifmmode~\acute{n}\else \'{n}\fi{}ski},
  \citenamefont {Chhajlany}, \citenamefont {Dutta}, \citenamefont
  {Tagliacozzo},\ and\ \citenamefont {Lewenstein}}]{2015SowinskiPRA}%
  \BibitemOpen
  \bibfield  {author} {\bibinfo {author} {\bibfnamefont {T.}~\bibnamefont
  {Sowi\ifmmode~\acute{n}\else \'{n}\fi{}ski}}, \bibinfo {author}
  {\bibfnamefont {R.~W.}\ \bibnamefont {Chhajlany}}, \bibinfo {author}
  {\bibfnamefont {O.}~\bibnamefont {Dutta}}, \bibinfo {author} {\bibfnamefont
  {L.}~\bibnamefont {Tagliacozzo}}, \ and\ \bibinfo {author} {\bibfnamefont
  {M.}~\bibnamefont {Lewenstein}},\ }\href {\doibase
  10.1103/PhysRevA.92.043615} {\bibfield  {journal} {\bibinfo  {journal} {Phys.
  Rev. A}\ }\textbf {\bibinfo {volume} {92}},\ \bibinfo {pages} {043615}
  (\bibinfo {year} {2015})}\BibitemShut {NoStop}%
\bibitem [{\citenamefont {Schliemann}\ \emph {et~al.}(2001)\citenamefont
  {Schliemann}, \citenamefont {Cirac}, \citenamefont {Ku\ifmmode~\acute{s}\else
  \'{s}\fi{}}, \citenamefont {Lewenstein},\ and\ \citenamefont
  {Loss}}]{2001SchliemannPRA}%
  \BibitemOpen
  \bibfield  {author} {\bibinfo {author} {\bibfnamefont {J.}~\bibnamefont
  {Schliemann}}, \bibinfo {author} {\bibfnamefont {J.~I.}\ \bibnamefont
  {Cirac}}, \bibinfo {author} {\bibfnamefont {M.}~\bibnamefont
  {Ku\ifmmode~\acute{s}\else \'{s}\fi{}}}, \bibinfo {author} {\bibfnamefont
  {M.}~\bibnamefont {Lewenstein}}, \ and\ \bibinfo {author} {\bibfnamefont
  {D.}~\bibnamefont {Loss}},\ }\href {\doibase 10.1103/PhysRevA.64.022303}
  {\bibfield  {journal} {\bibinfo  {journal} {Phys. Rev. A}\ }\textbf {\bibinfo
  {volume} {64}},\ \bibinfo {pages} {022303} (\bibinfo {year}
  {2001})}\BibitemShut {NoStop}%
\bibitem [{\citenamefont {Lindgren}\ \emph {et~al.}(2014)\citenamefont
  {Lindgren}, \citenamefont {Rotureau}, \citenamefont {Forss{\'{e}}n},
  \citenamefont {Volosniev},\ and\ \citenamefont {Zinner}}]{2014LindgrenNJP}%
  \BibitemOpen
  \bibfield  {author} {\bibinfo {author} {\bibfnamefont {E.~J.}\ \bibnamefont
  {Lindgren}}, \bibinfo {author} {\bibfnamefont {J.}~\bibnamefont {Rotureau}},
  \bibinfo {author} {\bibfnamefont {C.}~\bibnamefont {Forss{\'{e}}n}}, \bibinfo
  {author} {\bibfnamefont {A.~G.}\ \bibnamefont {Volosniev}}, \ and\ \bibinfo
  {author} {\bibfnamefont {N.~T.}\ \bibnamefont {Zinner}},\ }\href {\doibase
  10.1088/1367-2630/16/6/063003} {\bibfield  {journal} {\bibinfo  {journal}
  {New Journal of Physics}\ }\textbf {\bibinfo {volume} {16}},\ \bibinfo
  {pages} {063003} (\bibinfo {year} {2014})}\BibitemShut {NoStop}%
\bibitem [{\citenamefont {Chung}\ and\ \citenamefont
  {Bolech}(2017)}]{2017ChungPRA}%
  \BibitemOpen
  \bibfield  {author} {\bibinfo {author} {\bibfnamefont {S.~S.}\ \bibnamefont
  {Chung}}\ and\ \bibinfo {author} {\bibfnamefont {C.~J.}\ \bibnamefont
  {Bolech}},\ }\href {\doibase 10.1103/PhysRevA.96.023609} {\bibfield
  {journal} {\bibinfo  {journal} {Phys. Rev. A}\ }\textbf {\bibinfo {volume}
  {96}},\ \bibinfo {pages} {023609} (\bibinfo {year} {2017})}\BibitemShut
  {NoStop}%
\bibitem [{\citenamefont {Volosniev}(2017)}]{2017VolosnievFBS}%
  \BibitemOpen
  \bibfield  {author} {\bibinfo {author} {\bibfnamefont {A.~G.}\ \bibnamefont
  {Volosniev}},\ }\href {\doibase 10.1007/s00601-017-1227-0} {\bibfield
  {journal} {\bibinfo  {journal} {Few-Body Systems}\ }\textbf {\bibinfo
  {volume} {58}},\ \bibinfo {pages} {54} (\bibinfo {year} {2017})}\BibitemShut
  {NoStop}%
\bibitem [{\citenamefont {Cui}\ and\ \citenamefont {Ho}(2013)}]{2013CuiPRL}%
  \BibitemOpen
  \bibfield  {author} {\bibinfo {author} {\bibfnamefont {X.}~\bibnamefont
  {Cui}}\ and\ \bibinfo {author} {\bibfnamefont {T.-L.}\ \bibnamefont {Ho}},\
  }\href {\doibase 10.1103/PhysRevLett.110.165302} {\bibfield  {journal}
  {\bibinfo  {journal} {Phys. Rev. Lett.}\ }\textbf {\bibinfo {volume} {110}},\
  \bibinfo {pages} {165302} (\bibinfo {year} {2013})}\BibitemShut {NoStop}%
\bibitem [{\citenamefont {Kang}\ \emph {et~al.}(2018)\citenamefont {Kang},
  \citenamefont {Sun}, \citenamefont {Kang}, \citenamefont {Li},\ and\
  \citenamefont {Tan}}]{kang2018spontaneous}%
  \BibitemOpen
  \bibfield  {author} {\bibinfo {author} {\bibfnamefont {Y.}~\bibnamefont
  {Kang}}, \bibinfo {author} {\bibfnamefont {Z.}~\bibnamefont {Sun}}, \bibinfo
  {author} {\bibfnamefont {Y.}~\bibnamefont {Kang}}, \bibinfo {author}
  {\bibfnamefont {Y.}~\bibnamefont {Li}}, \ and\ \bibinfo {author}
  {\bibfnamefont {S.}~\bibnamefont {Tan}},\ }\href
  {https://doi.org/10.1007/s10909-017-1833-8} {\bibfield  {journal} {\bibinfo
  {journal} {Journal of Low Temperature Physics}\ }\textbf {\bibinfo {volume}
  {190}},\ \bibinfo {pages} {225} (\bibinfo {year} {2018})}\BibitemShut
  {NoStop}%
\end{thebibliography}%

\newpage
\renewcommand{\thefigure}{S\arabic{figure}}
\setcounter{figure}{0}
\setcounter{page}{1}
\onecolumngrid

\begin{center}
{\bf \large Geometry-induced entanglement in a mass-imbalanced few-fermion system \\[5pt] SUPPLEMENTARY MATERIAL}\\[10pt]

Damian W{\l}odzy\'nski$^{(1)}$, Daniel P{\c e}cak$^{(2)}$, and Tomasz Sowi\'nski$^{(1)}$ \\[3pt]

\mbox{\it $^{(1)}$Institute of Physics, Polish Academy of Sciences, Aleja Lotnik\'ow 32/46, PL-02668 Warsaw, Poland} \\
\mbox{\it $^{(2)}$Faculty of Physics, Warsaw University of Technology, Ulica Koszykowa 75, PL-00662 Warsaw, Poland}
\end{center}

\begin{figure}[h!]
\centering
\includegraphics[width=0.45\linewidth]{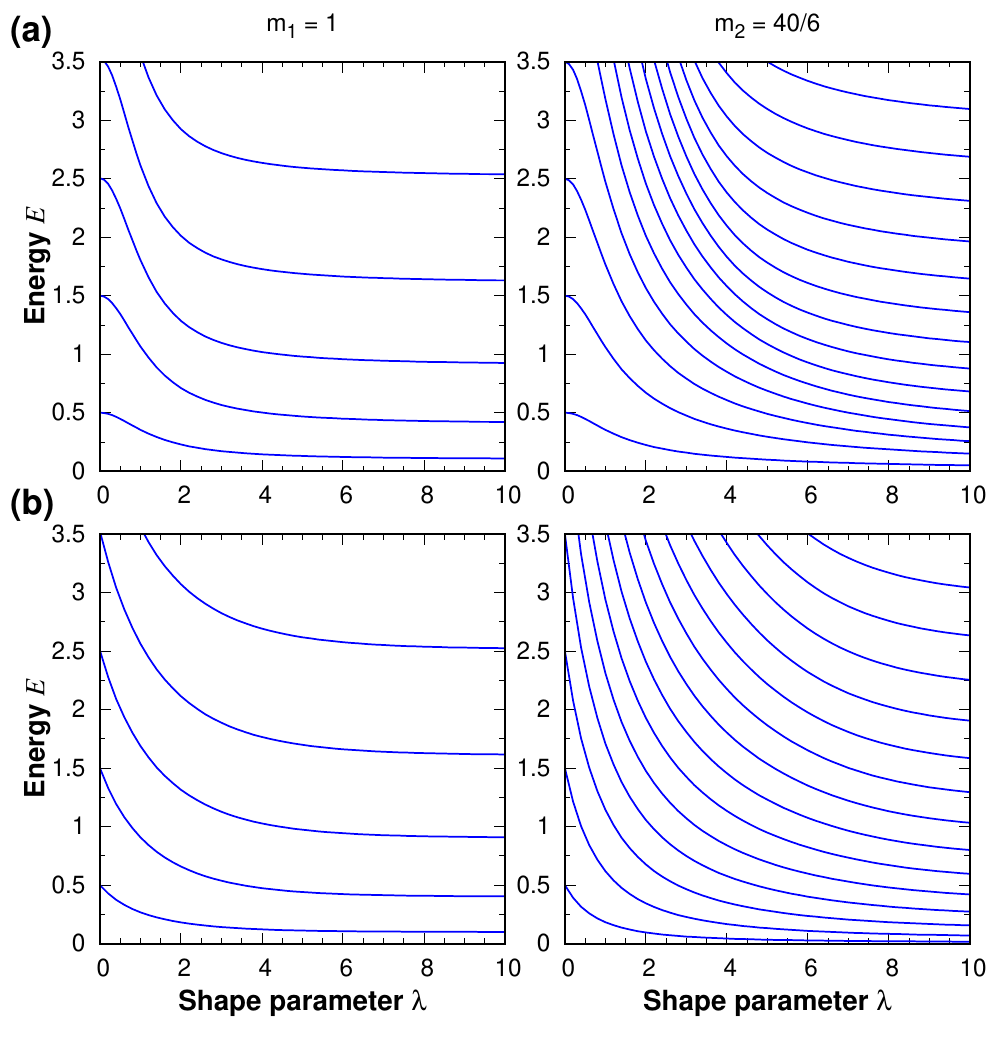}
\caption{The single-particle energy spectrum $E$ in function of the shape parameter $\lambda$ for different masses ($m_1=1$ and $m_2=40/6$). Transition from harmonic confinement ($\lambda=0$) to uniform confinement ($\lambda \to \infty$) is described by function: (a) $f_1(\lambda)$, (b) $f_2(\lambda)$.  \label{FigS1} }
\end{figure}

\begin{figure}
\centering
\includegraphics[width=0.45\linewidth]{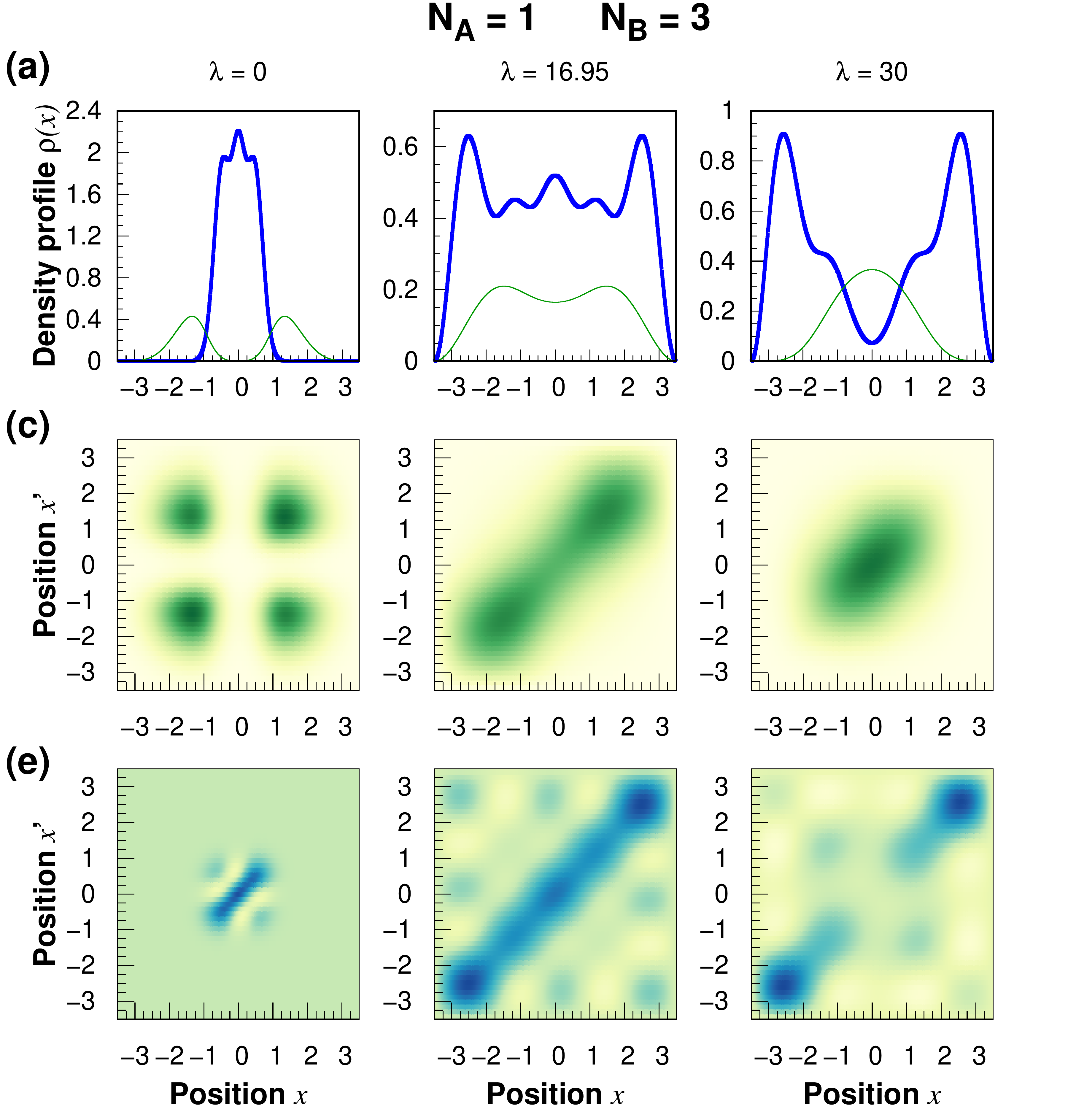}
\includegraphics[width=0.45\linewidth]{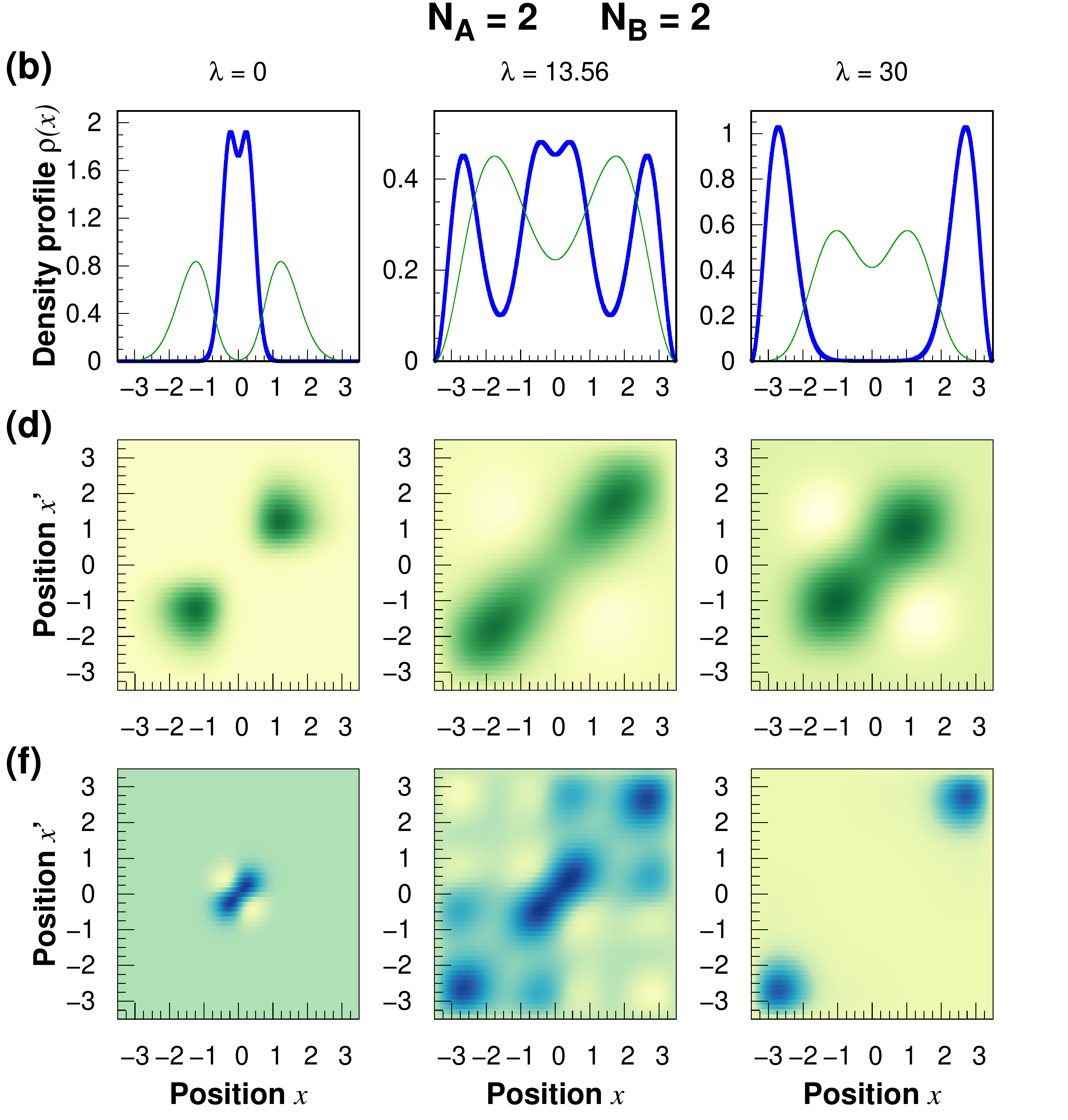}
\caption{Structural transition in the ground state of the system of $N_A+N_B=4$ fermions (for interaction strength $g=5$ and mass ratio  $m_B/m_A=40/6$). Successive columns (from left to right) correspond to different external traps, from the harmonic oscillator ($\lambda=0$) to the flat box ($\lambda\rightarrow\infty$). (a-b) Single-particle density profile for heavier (thick blue) and lighter (thin green) component depending on the shape of the external trap. (c-d) Single-particle density matrix of the lighter component. (e-f) Single-particle density matrix of the heavier component.\label{FigS2}}
\end{figure}

\begin{figure}[t]
\centering
\includegraphics[width=0.45\linewidth]{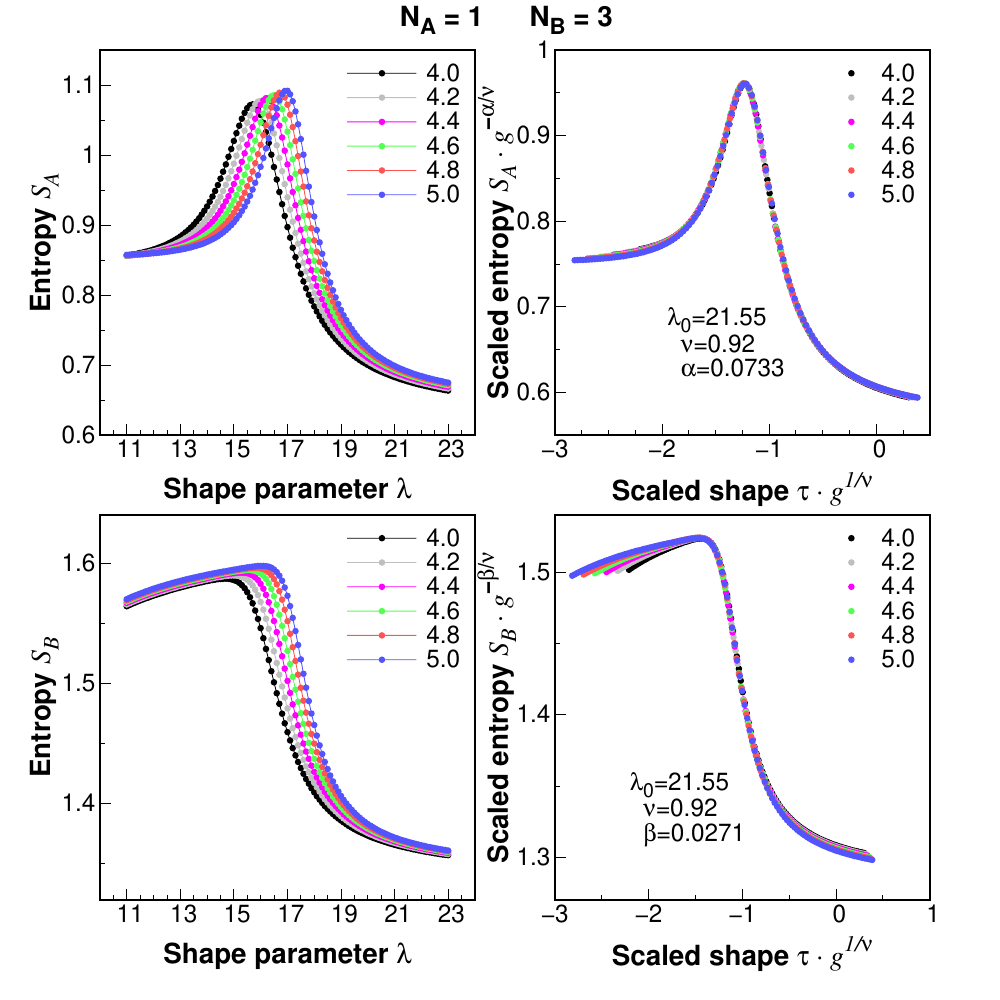}
\includegraphics[width=0.45\linewidth]{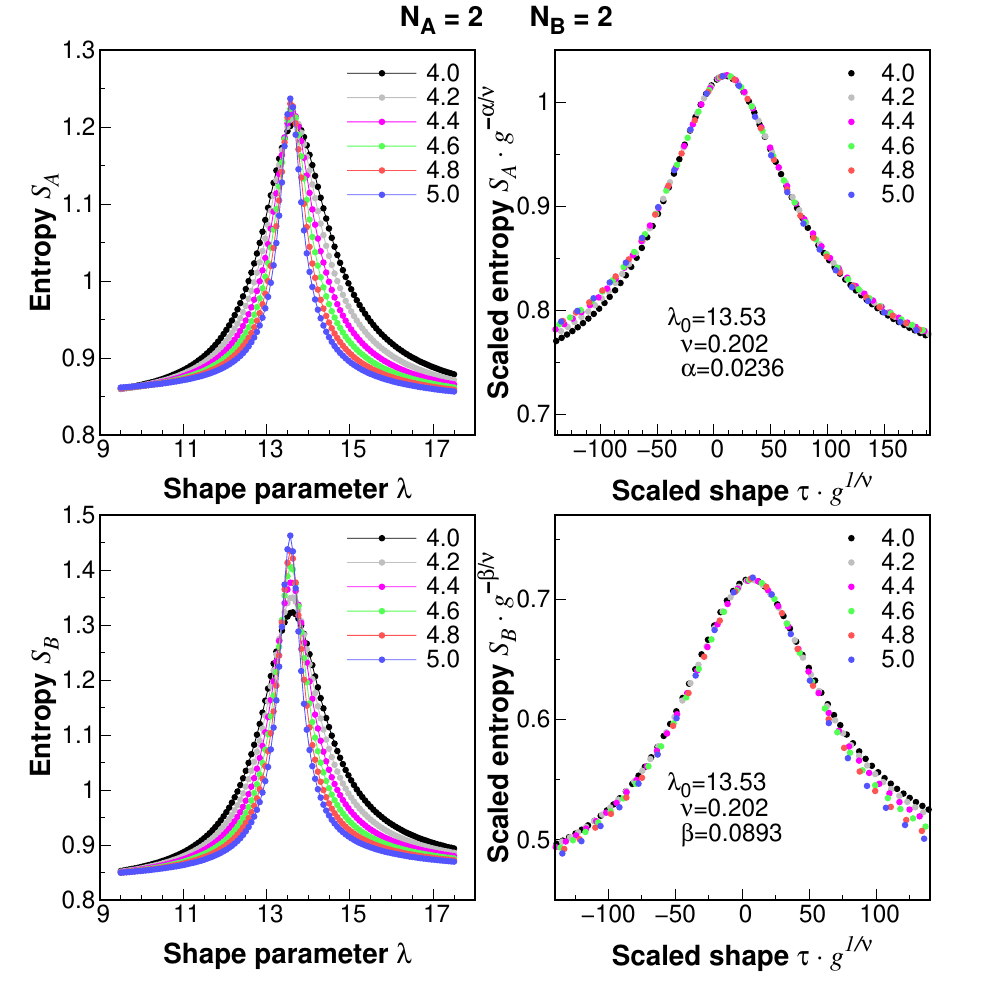}
\caption{The single-particle von Neumann entropies $S_A$ and $S_B$ as a function of shape parameter $\lambda$ and interaction strength $g$ calculated for different distributions of $N = 4$ particles. After appropriate scaling, data for different interaction strength g collapse to single universal curve.\label{FigS3}}
\end{figure}

\begin{figure}[t]
\centering
\includegraphics[width=0.45\linewidth]{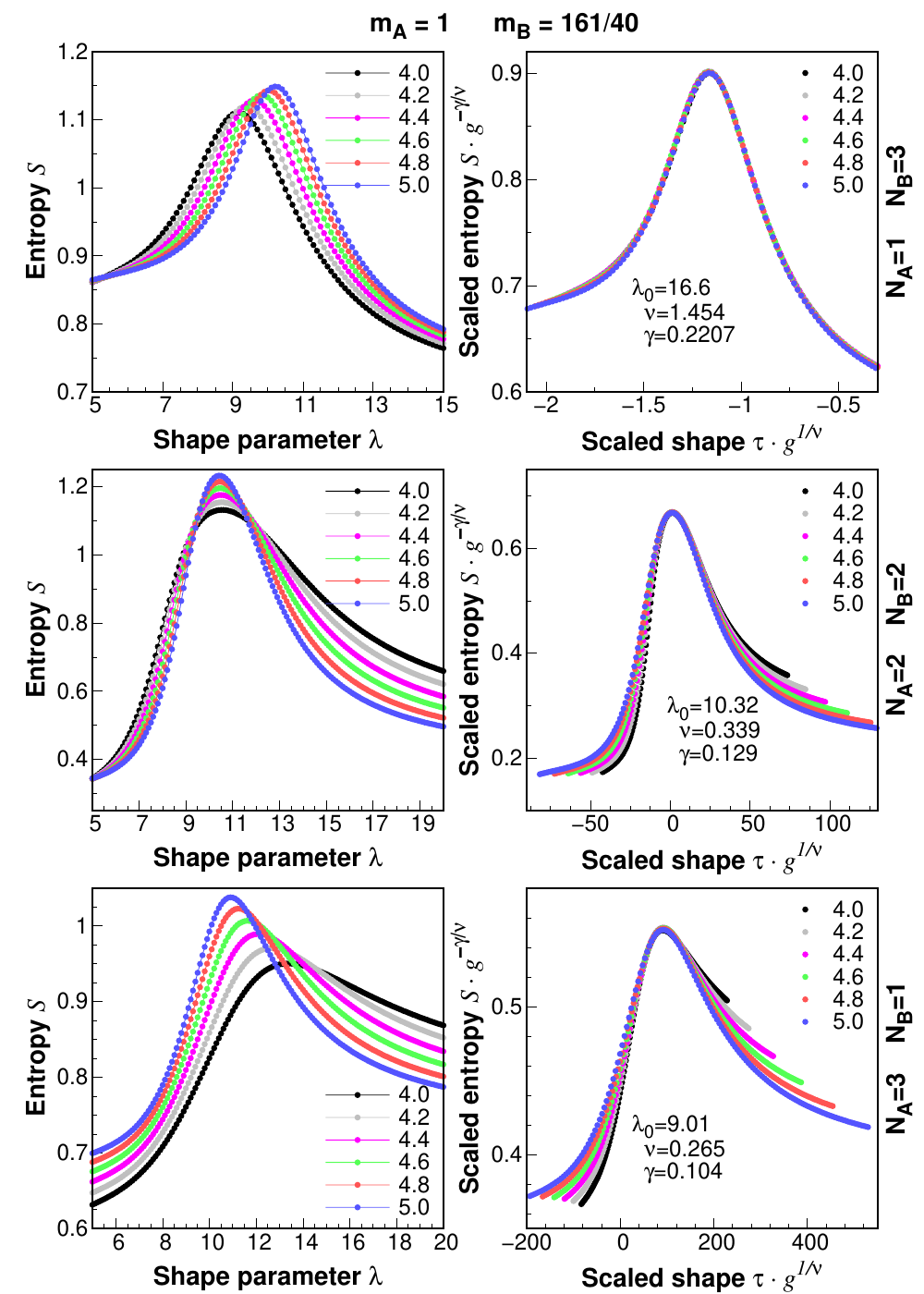}
\includegraphics[width=0.45\linewidth]{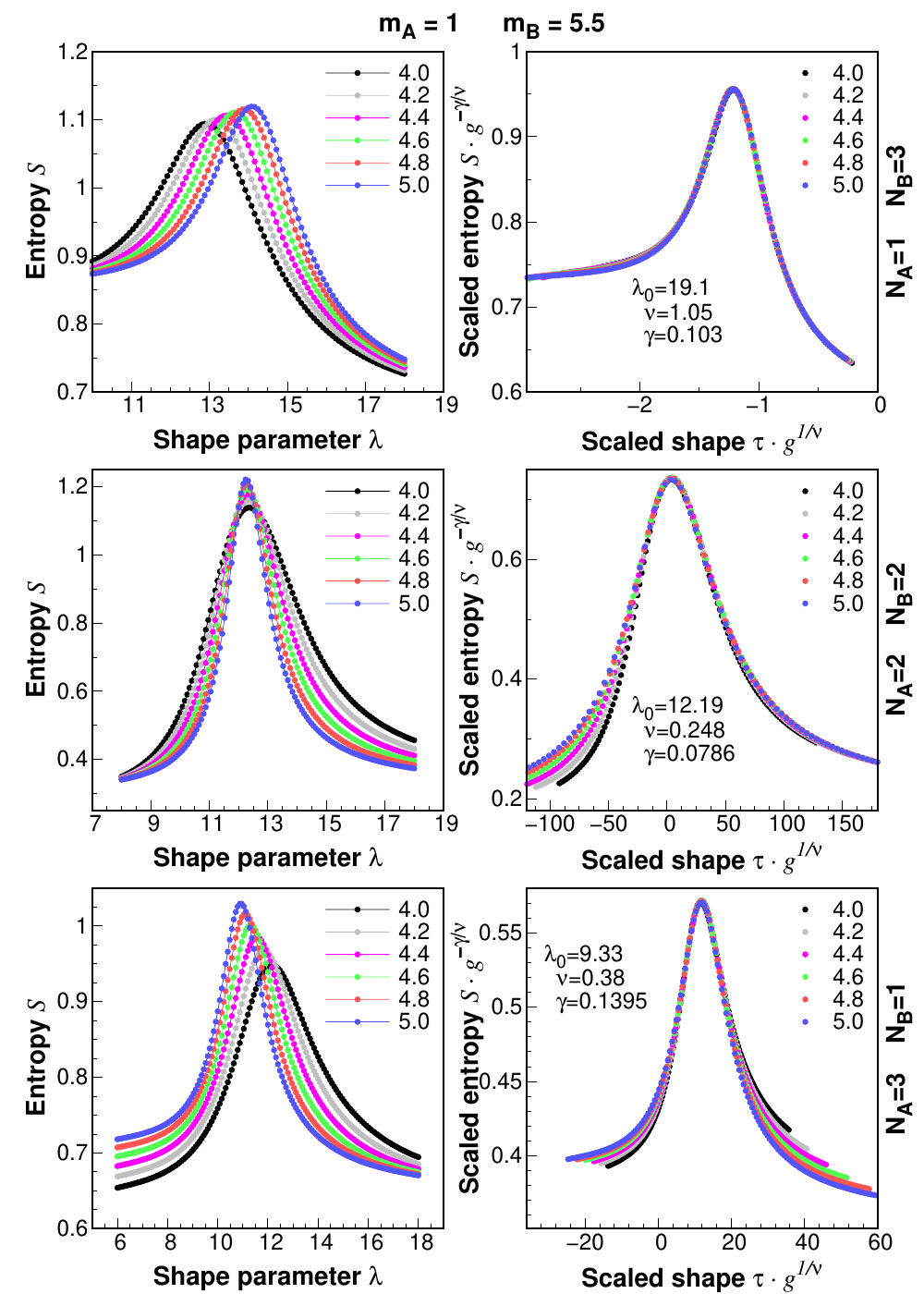}
\caption{The inter-component entanglement entropy as a function of a shape of external trap $\lambda$ and interaction strength g calculated for different distributions of $N = 4$ particles and for two different mass ratios: $m_B/m_A = 161/40$ (corresponding to the dysprosium-potassium mixture) and $m_B/m_A = 5.5$. After appropriate scaling, data for different interaction strength $g$ collapse to a single universal curve.\label{FigS4}}
\end{figure}

\begin{figure}[t]
\centering
\includegraphics[width=1.05\linewidth]{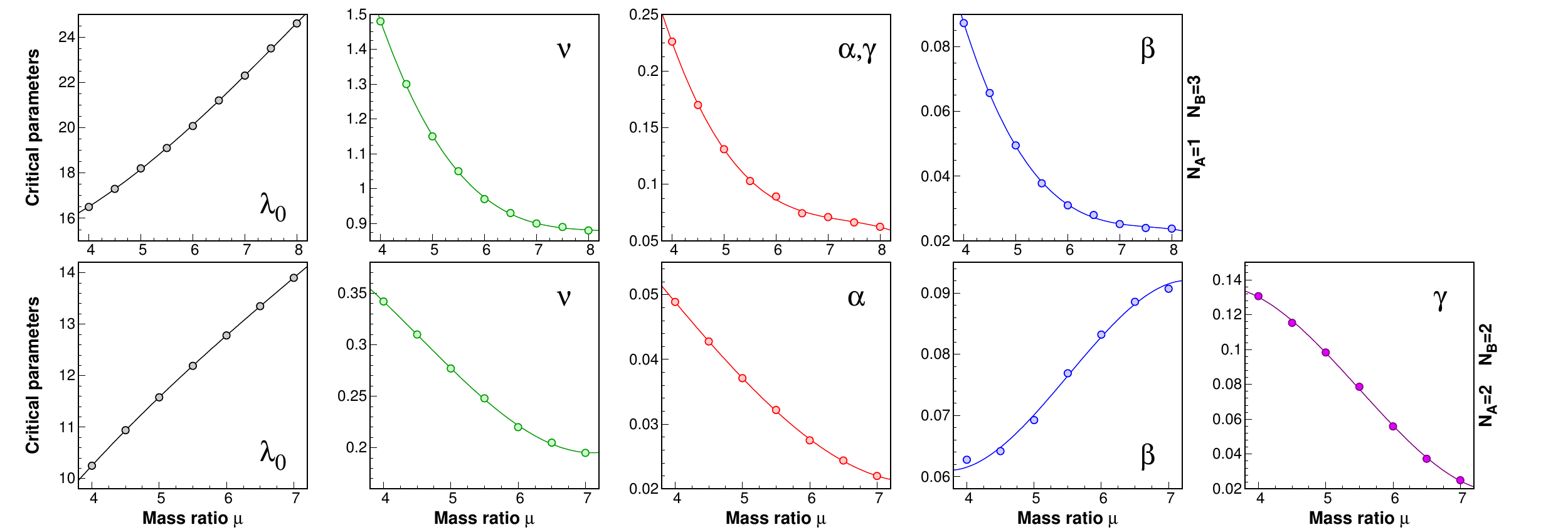}
\caption{Values of the critical parameters $\lambda_0$, $\alpha$, $\beta$, $\gamma$, and $\nu$ as functions of the mass ratio $\mu$ for the systems with $(N_A,N_B) = (1,3)$ and $(N_A,N_B) = (2,2)$ fermions. It is clear that all the parameters are significantly and monotonically dependent on the ratio. 
 \label{FigS5}}
\end{figure}

\begin{figure}[t]
\centering
\includegraphics[width=0.45\linewidth]{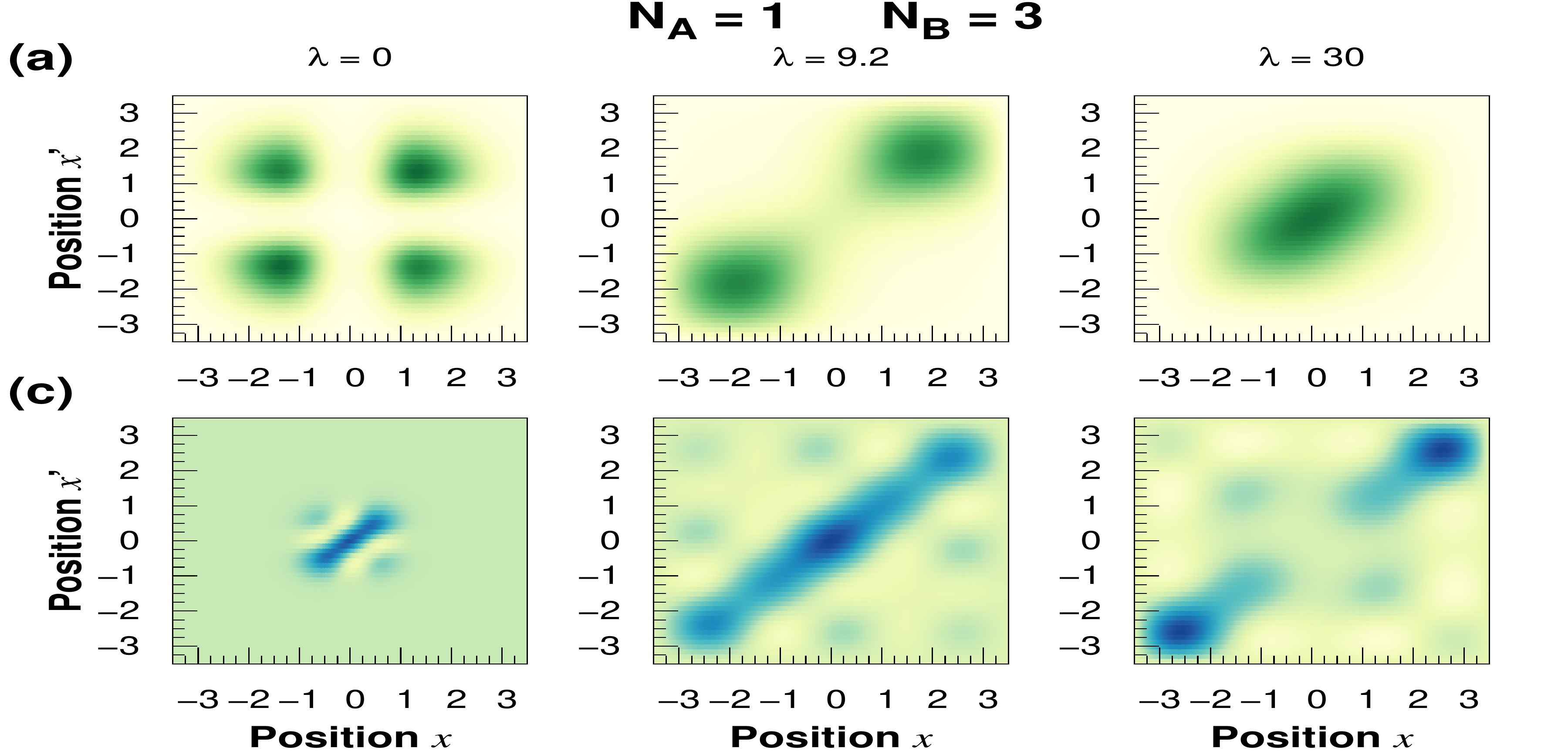}
\includegraphics[width=0.45\linewidth]{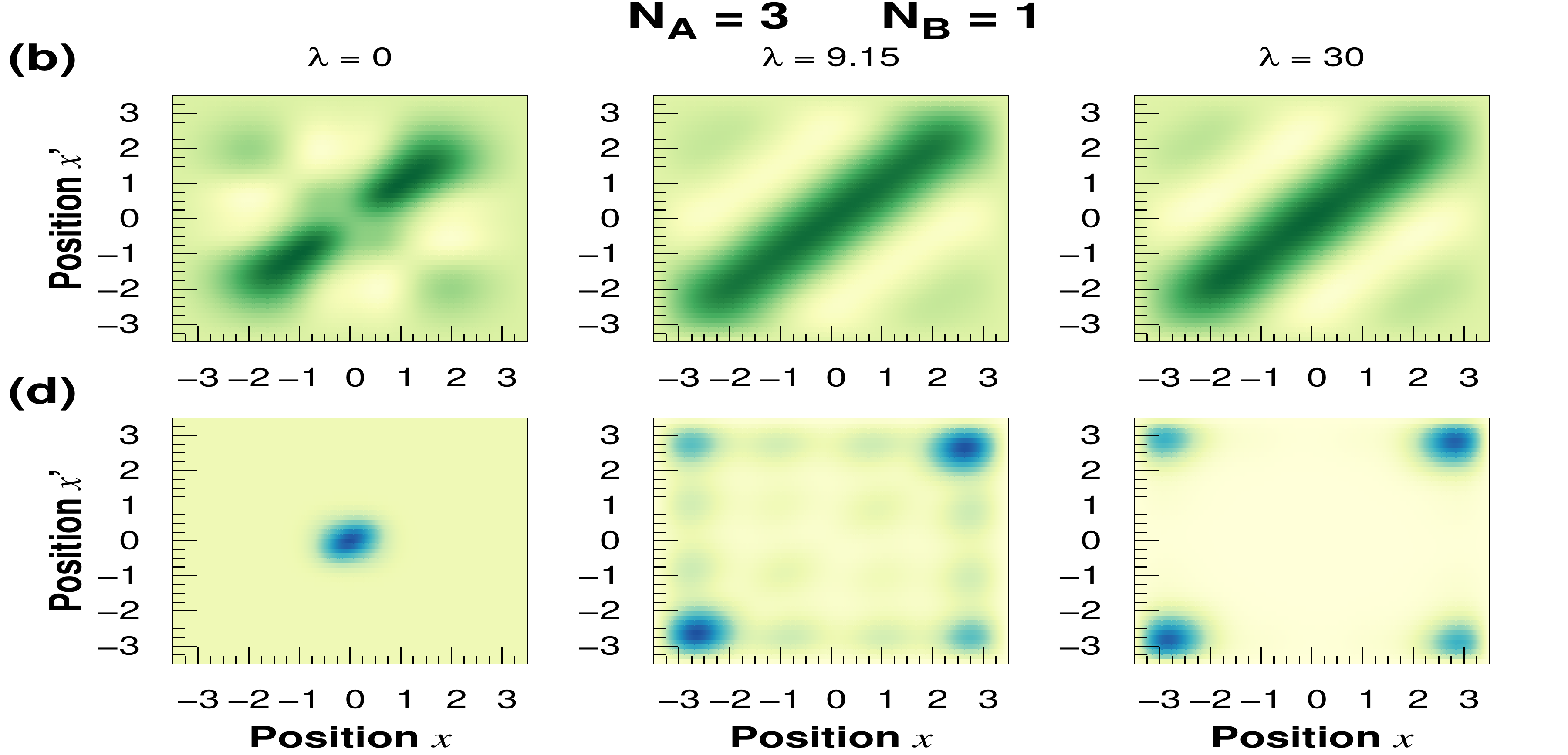}
\includegraphics[width=0.45\linewidth]{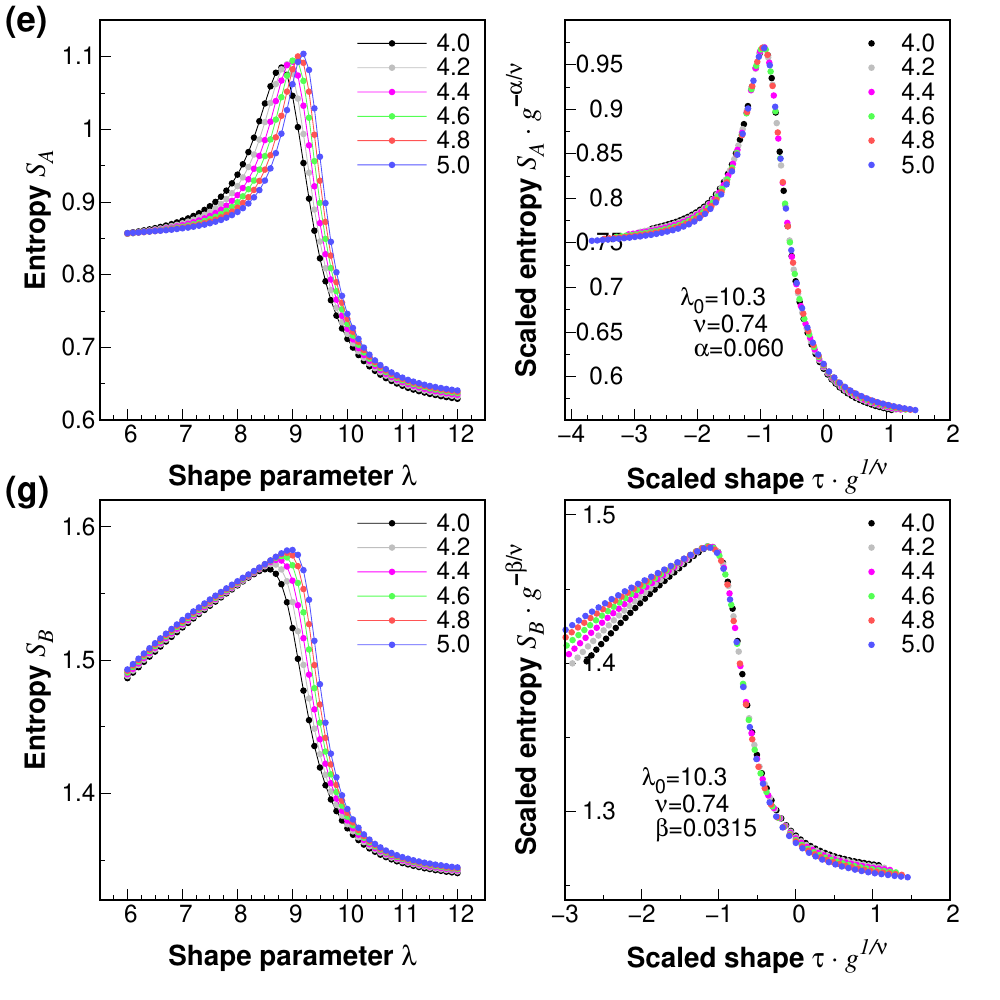}
\includegraphics[width=0.45\linewidth]{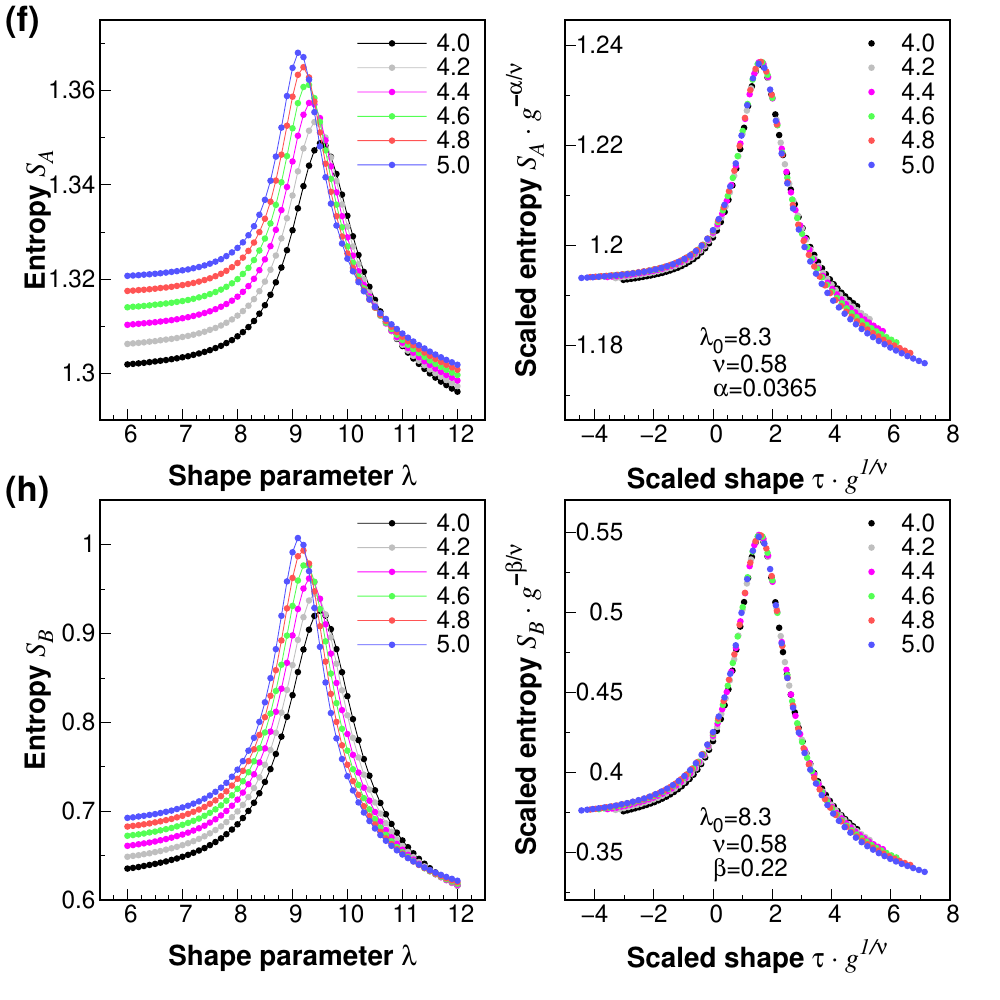}
\caption{Structural transition in the ground state of the system of $N_A+N_B=4$ fermions ($m_B/m_A=40/6$), with confinement's shape described by function $f_2(\lambda)$. (a-d) Single-particle density matrix of the lighter and heavier component (first and second row respectively), for interaction strength $g=5$. Successive columns (from left to right) correspond to different external traps, from the harmonic oscillator ($\lambda=0$) to the flat box ($\lambda\rightarrow\infty$). (e-h) The single-particle von Neumann entropies $S_A$ and $S_B$ (third and fourth row respectively) as a function of the shape parameter $\lambda$ and interaction strength $g$, with corresponding universal curves. \label{FigS6}}
\end{figure}

\begin{figure}[t]
\centering
\includegraphics[width=0.45\linewidth]{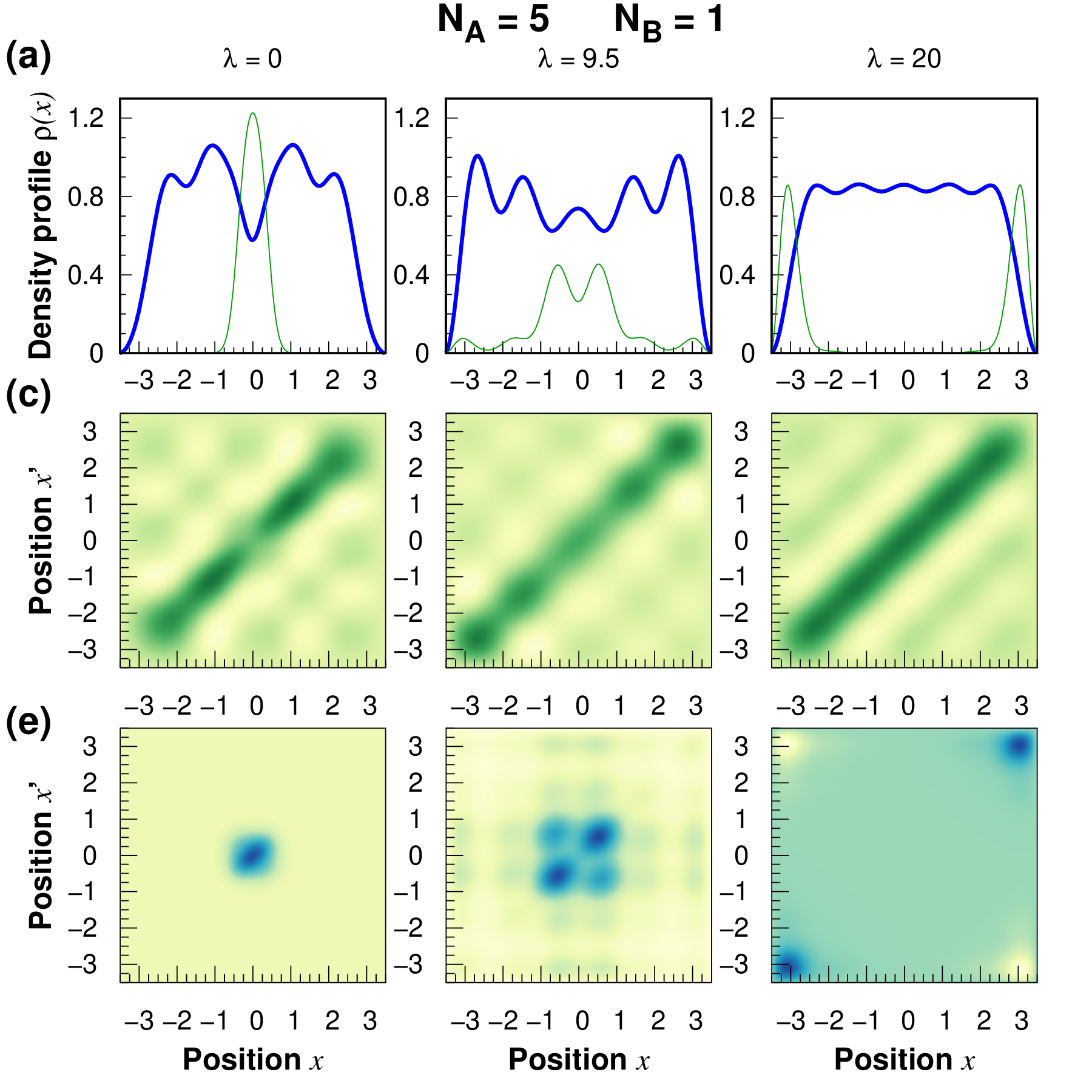}
\includegraphics[width=0.45\linewidth]{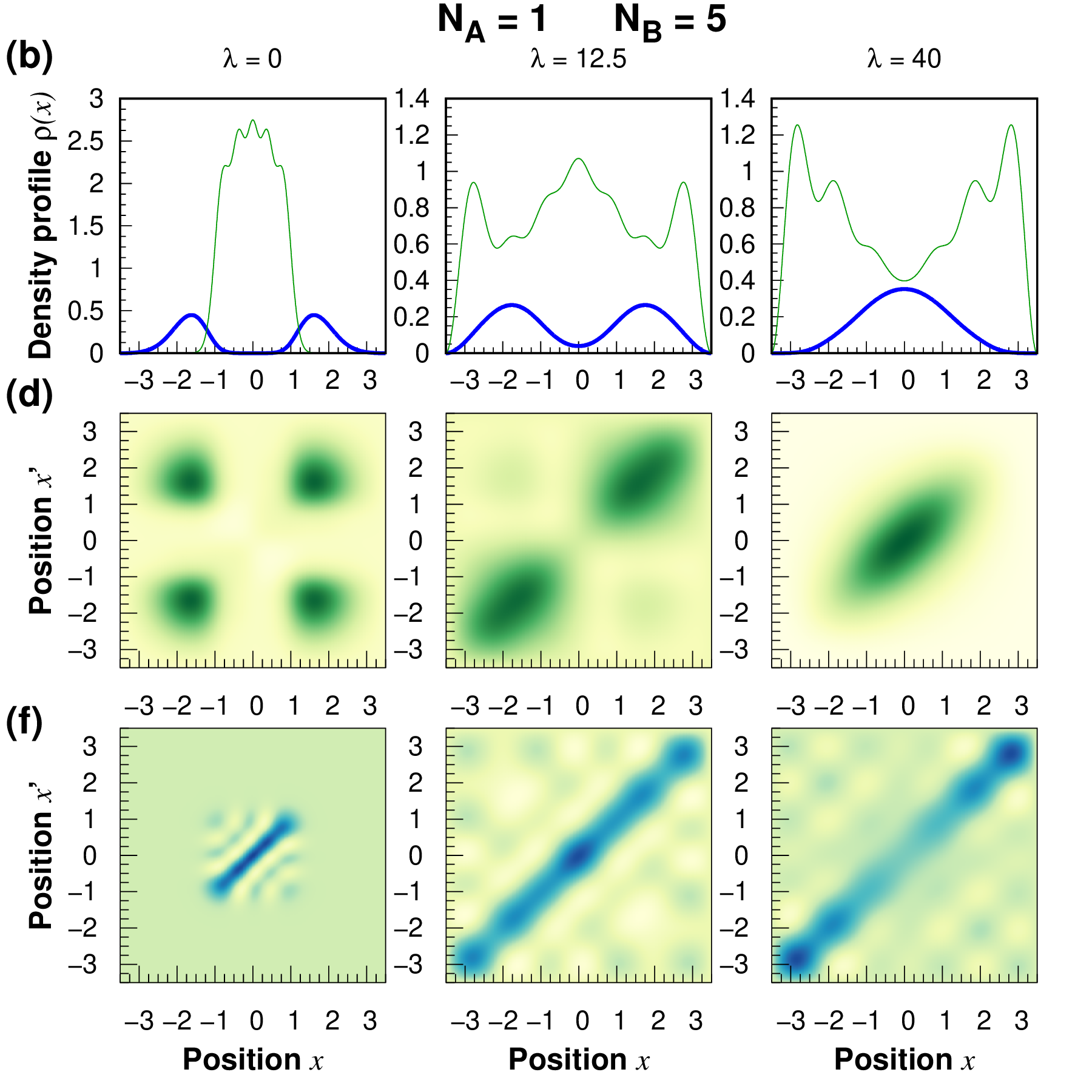}
\includegraphics[width=0.45\linewidth]{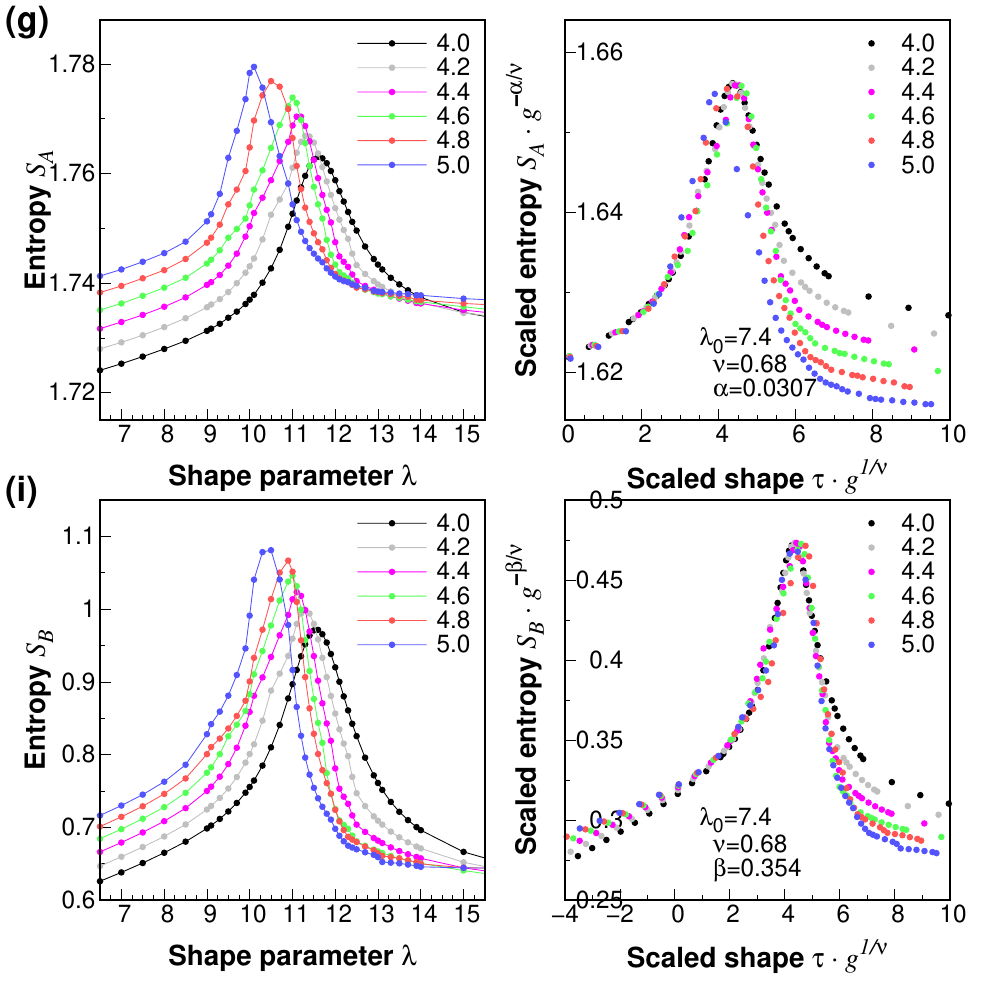}
\includegraphics[width=0.45\linewidth]{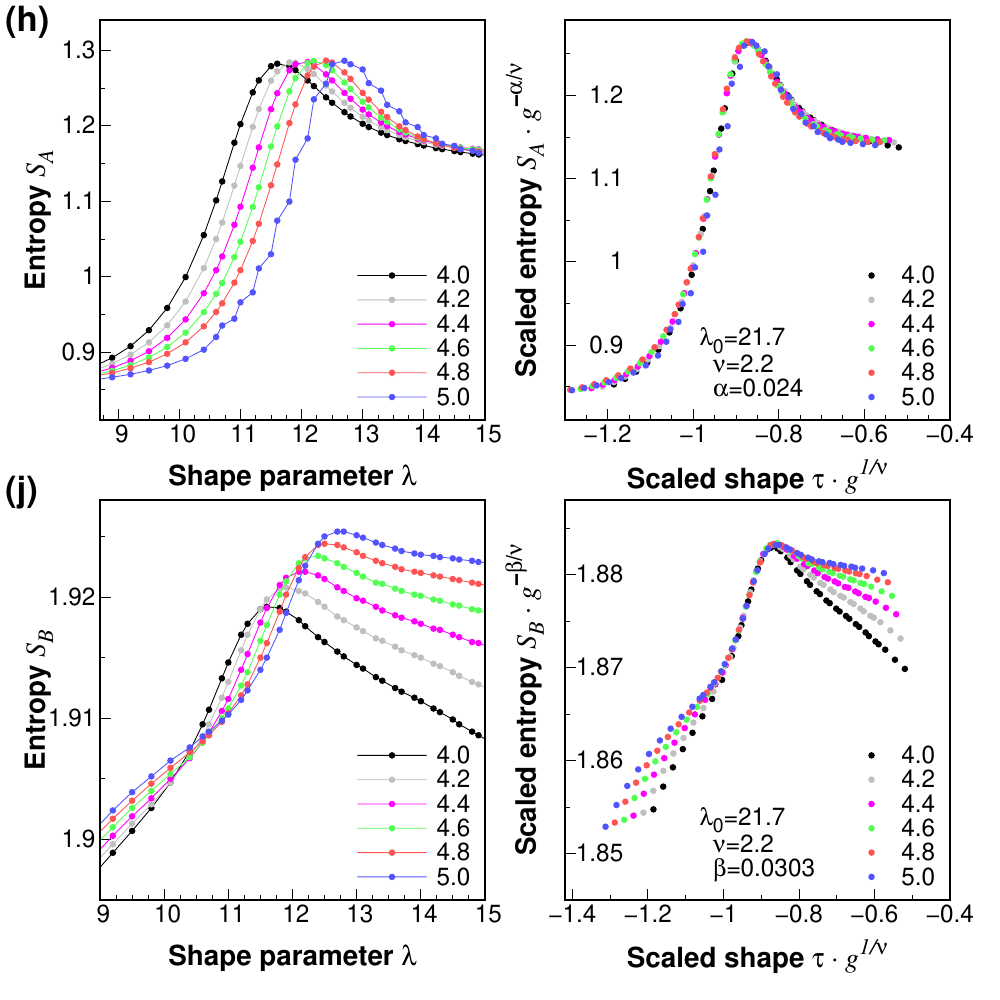}
\caption{Structural transition in the ground state of system with $N_A+N_B=6$ fermions ($m_B/m_A=40/6$) and imbalance $|N_A-N_B|=4$ in a trap with a shape described by the function $f_2(\lambda)$. (a-b) Single-particle density profile for heavier (thick blue) and lighter (thin green) component depending on the shape of the external trap and interaction strength $g=5$. (c-f) Single-particle density matrix of the lighter and heavier component (first and second row respectively). Successive columns (from left to right) correspond to different external traps, from the harmonic oscillator ($\lambda=0$) to the flat box ($\lambda\rightarrow\infty$). (g-j) The single-particle von Neumann entropies $S_A$ and $S_B$ (third and fourth row respectively) as a function of the shape parameter $\lambda$ and interaction strength $g$, with corresponding universal curves. Note, that considered interactions are still to weak to lead to a perfect separation. Due to numerical limitations it is not possible to obtain results for stronger interactions. However, characteristic scaling is clearly visible.\label{FigS7}}
\end{figure}

\end{document}